\documentclass[11pt]{article}
\usepackage{amssymb}
\usepackage{graphics}
\usepackage{epsfig}
\usepackage{a4wide}
\usepackage{slashed}

\begin{document}

\newcommand{\ppaq}{$pp \to A_H q_- + X~$ }
\newcommand{\qgaq}{$qg \to A_H q_-~$}
\newcommand{\ugau}{$ug \to A_H u_-~$}
\newcommand{\qgaqg}{$qg \to A_H q_-+g~$}
\newcommand{\ggaqq}{$gg \to A_Hq_-+\bar{q}~$}
\newcommand{\qqaqq}{$qq' \to A_H q_- + q~$}
\newcommand{\uuauu}{$q\bar{q}\to A_H q_-+\bar{q}$}
\title{  Precise predictions for $A_H q_-$ associated production
         in the littlest Higgs model with $T$ parity at the LHC }
\author{ Yang Xiao-Dong, Xiong Shou-Jian, Ma Wen-Gan, Zhang Ren-You, Guo Lei, and Li Xiao-Zhou \\
{\small  Department of Modern Physics, University of Science and Technology}  \\
{\small  of China (USTC), Hefei, Anhui 230026, P.R.China}}

\date{}
\maketitle \vskip 15mm
\begin{abstract}
In the framework of the littlest Higgs model with $T$ parity, we
present complete calculations for the $A_H q_-$ $(q_-=u_-,
\bar{u}_-,d_-, \bar{d}_-,c_-, \bar{c}_-,s_-, \bar{s}_-)$ associated
production up to the QCD next-to-leading order (NLO) at the CERN
Large Hadron Collider with subsequent pure weak decay of $T$-odd
mirror quark. We apply the PROSPINO scheme to avoid the double counting
problem and to keep the convergence of the perturbative QCD
description. The theoretical correlations between the integrated
cross section and the factorization and renormalization scale, the
global symmetry-breaking scale and the Yukawa coupling parameter are
studied separately. We also provide the kinematic distributions of
the final decay products. Our numerical results show that the NLO
QCD correction reduces the scale uncertainty and enhances the leading-order
integrated cross section remarkably, with the $K$ factor varying in
the range of $1.41 \sim 1.68$ ($1.58 \sim 1.89$) as the increment of
the global symmetry-breaking scale $f$ from $500~{\rm GeV}$ to
$1.5~{\rm TeV}$ ($1.1~{\rm TeV}$) at the $\sqrt{s} = 14~{\rm TeV}$
($8~{\rm TeV}$) LHC. We find that it is possible to select the signal
events of the $A_Hq_-$ production from its background by putting proper
cuts on the final leading jet and missing energy.
\end{abstract}

\vskip 15mm {\large\bf PACS: 12.38.Bx, 12.60.Cn, 14.70.Pw  }

\vfill \eject \baselineskip=0.32in

\renewcommand{\theequation}{\arabic{section}.\arabic{equation}}
\renewcommand{\thesection}{\Roman{section}.}
\newcommand{\nb}{\nonumber}

%slash:
\newcommand{\Dir}{\kern -6.4pt\Big{/}}%su lettere italiane minuscole
\newcommand{\Dirin}{\kern -10.4pt\Big{/}\kern 4.4pt}
\newcommand{\DDir}{\kern -7.6pt\Big{/}}%su lettere italiane maiuscole
\newcommand{\DGir}{\kern -6.0pt\Big{/}}%su lettere greche

\makeatletter      % '@' is now a normal "letter" for TeX
\@addtoreset{equation}{section}
\makeatother       % '@' is restored as a "non-letter" character for TeX

\par
\section{Introduction }
\par
The standard model (SM) \cite{s1,s2} provides an excellent
description of high-energy phenomena at the energy scale up to
$10^2~{\rm GeV}$. However, several theoretical problems
\cite{Barbieri:2000gf} that the SM encounters still make us
confused, and drive physicists to consider new physics beyond the
SM. Many extensions of the SM are proposed to deal with these
problems, such as supersymmetric models \cite{model-2}, extra
dimensions models \cite{model-3}, little Higgs models
\cite{LittleHiggs}, grand unified theories \cite{model-1}, and so on.
Among them, the little Higgs models deserve much attention due to
their elegant solution to the hierarchy problem, and are proposed as
one kind of electroweak symmetry-breaking (EWSB) model without
fine tuning, in which the Higgs boson is naturally light as a result
of nonlinearly realized symmetry \cite{Arkani,LH7}. The simplest
version of the little Higgs models is the littlest Higgs model (LH)
\cite{LH4}, which is based on an $SU(5)/SO(5)$ nonlinear $\sigma$
model. In the LH, a set of new heavy gauge bosons
($A_H,Z_H,W_H^{\pm}$) and a vectorlike quark ($T$) are introduced
to cancel the quadratic divergence of the Higgs boson mass contributed
by the SM gauge boson loops and the top quark loop, respectively.
However, the present precision electroweak constraints
\cite{EWC,Limit-LHT-1,Limit-LHT-2} require the LH to be characterized
by a large value of the global symmetry-breaking scale $f$, so the
fine tuning between the cutoff scale and the electroweak scale is
again needed. Fortunately, this problem can be solved by introducing
a discrete symmetry, the $T$ parity, to the LH
\cite{Low:2004xc,Cheng:2003ju}.

\par
In the littlest Higgs model with $T$ parity (LHT), all the SM
particles are $T$-even and all the new heavy particles except $T_+$
are $T$ odd. Then the SM gauge bosons cannot mix
with the new heavy gauge bosons, and the global symmetry-breaking
scale $f$ can be lower than $1~{\rm TeV}$ \cite{Hubisz:2005tx}.
Recently, from the analyses of Higgs data from the ATLAS and CMS
Collaborations combined with other experimental results, we get the
constraint on the scale $f$ in the LHT as $f > 694~{\rm GeV}$ given in
Ref.\cite{Limit-LHT-1} and $f > 665~{\rm GeV}$ provided in
Ref.\cite{Limit-LHT-2}. In the LHT, the decay channels
$W_H^{\mp}\to l^{\mp}\stackrel{(-)}{\nu}$ and $Z_H \to l^{+}l^{-}$
are forbidden, and the probable decay modes of these heavy gauge
bosons should be $T$-parity conserving. What's more, the LHT offers
a candidate for dark matter, called the heavy photon $A_H$, which
cannot decay into other particles. As more and more attention is
paid to dark matter \cite{Asano}, the precision investigation for
$A_H$ production will be meaningful and necessary.

\par
In this paper, we focus on the $A_H q_-$ production up to the QCD NLO,
where $q_-$ represents the $T$-odd mirror quark of the first two
generations. A brief review of the related LHT theory can be found
in Sec.II. In Sec.III, we provide our calculation
strategy. The numerical results and discussions are presented in
Sec.IV. Finally a short summary is given.

\vskip 5mm
\section{Related LHT theory}\label{theory}
\par
In this section, we will briefly review the LHT theory related to
our calculations. For more details, one can refer to
Refs.\cite{cpyuan:2006ph,Blanke,sasha-pheno}.

\par
The LH is based on a nonlinear $\sigma$ model describing the spontaneous breaking
of an $SU(5)$ global symmetry down to its $SO(5)$ subgroup at an energy scale $f \sim 1~{\rm TeV}$.
This symmetry breaking originates from the vacuum expectation value (VEV) of an
$SU(5)$ symmetric tensor field $\Sigma$, given by
\begin{eqnarray}
\Sigma_0 = \langle \Sigma \rangle = \left(
 \begin{array}{ccccc}  & & 1_{2 \times 2} \\  & 1 & \\  1_{2 \times 2} & & \end{array}
 \right).
\end{eqnarray}
Then the nonlinear $\sigma$ model tensor field $\Sigma$ can be written as
\begin{eqnarray}
\Sigma = e^{i \Pi/f} \Sigma_0 e^{i \Pi^{T}/f} = e^{2 i \Pi/f} \Sigma_0,
\end{eqnarray}
where $\Pi(x)$ is the ``pion'' matrix containing the 14 Nambu-Goldstone degrees of freedom from the
$SU(5)/SO(5)$ breaking.
An $[SU(2)\times U(1)]_{1}\times[SU(2)\times U(1)]_{2}$ subgroup of the
$SU(5)$ global symmetry is gauged in the LH, and the gauge
fields $W_{i \mu}^a$ and $B_{i\mu}$ $(a = 1, 2, 3,~ i = 1, 2)$ are introduced.
To implement $T$ parity in the LHT, we make the following $T$ parity assignment:
\begin{eqnarray}
 W_{1\mu}^a \longleftrightarrow W_{2\mu}^a,~~~~
 B_{1\mu} \longleftrightarrow B_{2\mu},~~~~
\Pi \longrightarrow -\Omega \Pi \Omega,
\end{eqnarray}
where $\Omega = {\rm diag}(1,1,-1,1,1)$. Due to T-parity conservation, the
gauge couplings of the two $SU(2)\times U(1)$ subgroups
have to be equal, i.e., $g_1 = g_2 = \sqrt{2} g$ and $g_1^{\prime} =
g_2^{\prime} = \sqrt{2} g^{\prime}$. The $T$-odd and $T$-even gauge fields
can be obtained as
\begin{eqnarray}
 W_L^a &=& \frac{W_1^a + W_2^a}{\sqrt{2}},~~~
 B_L = \frac{B_1 + B_2}{\sqrt{2}},~~~~~ (\mbox{$T$ even}), \nonumber \\
 W_H^a &=& \frac{W_1^a - W_2^a}{\sqrt{2}},~~~
 B_H = \frac{B_1 - B_2}{\sqrt{2}},~~~~~ (\mbox{$T$ odd}).
\end{eqnarray}
The VEV $\Sigma_0$ breaks the gauge group $[SU(2)\times
U(1)]_{1}\times[SU(2)\times U(1)]_{2}$ down to its diagonal
subgroup, which is identified with the SM electroweak gauge group
$SU(2)_L \times U(1)_Y$, and the electroweak symmetry breaking
(EWSB) $SU(2)_L \times U(1)_Y \to U(1)_{em}$ takes place via the
usual Higgs mechanism. The mass eigenstates of the gauge sector in
the LHT are given by
\begin{eqnarray}
&& W_L^{\pm} = \frac{W_L^1 \mp i W_L^2}{\sqrt{2}},~~~~
  \left( \begin{array}{c} A_L \\ Z_L \end{array} \right)
 = \left( \begin{array}{rc} \cos\theta_W & \sin\theta_W \\
 -\sin\theta_W & \cos\theta_W \end{array} \right)  \left(
 \begin{array}{c} B_L \\ W_L^3 \end{array} \right),~~~~(\mbox{$T$ even}), \nonumber \\
&& W_H^{\pm} = \frac{W_H^1 \mp i W_H^2}{\sqrt{2}}, ~~~
  \left( \begin{array}{c} A_H \\ Z_H \end{array} \right)
 = \left( \begin{array}{cr} \cos\theta_H & -\sin\theta_H \\
 \sin\theta_H & \cos\theta_H \end{array} \right)
 \left( \begin{array}{c} B_H \\ W_H^3 \end{array} \right),~~~~(\mbox{$T$ odd}),
\end{eqnarray}
where $\theta_{W}$ is the Weinberg angle and the mixing angle
$\theta_H$ at the ${\cal O}(v_{SM}^2/f^2)$ is defined as
\begin{eqnarray}
\sin{\theta_{H}} \simeq \frac{5 g g^{\prime}}{4(5 g^2 - g^{\prime
2})} \frac{v_{SM}^2}{f^2}.
\end{eqnarray}
The $T$-even gauge bosons $A_L$, $Z_L$ and $W_L$ are identified with
the SM photon, $Z$ boson and $W$ boson, respectively, while the four new heavy gauge bosons are $A_H$,
$Z_H$, $W_H^{\pm}$. The $T$ odd gauge boson masses are given by
\begin{eqnarray}\label{mass-AH-VH}
 m_{A_H} \simeq \frac{1}{\sqrt{5}} g^{\prime} f
 \left( 1 - \frac{5}{8}\frac{v_{SM}^2}{f^2} \right),~~~~
 m_{W_H} \simeq g f \left( 1 - \frac{1}{8}\frac{v_{SM}^2}{f^2}
 \right),~~~~
 m_{Z_H} \simeq m_{W_H}.
\end{eqnarray}

\par
To implement $T$ parity in the quark sector we introduce two
incomplete $SU(5)$ multiplets and an $SO(5)$ multiplet
\footnote{Here we only consider one generation for demonstration purpose.},
\begin{eqnarray}
 \Psi_1 = \left( \begin{array}{c} \psi_1 \\ 0 \\ 0 \end{array}
 \right),~~~~  \Psi_2 = \left( \begin{array}{c} 0 \\ 0 \\ \psi_2 \end{array}
 \right),~~~~  \Psi_{HR} = \left( \begin{array}{c} \tilde{\psi}_{HR} \\
 \chi_{HR} \\  \psi_{HR} \end{array} \right),
\end{eqnarray}
with
\begin{eqnarray}
 \psi_i = -\tau^2 q_i = -\tau^2  \left( \begin{array}{c}
 u_i \\ d_i \end{array} \right),~~(i = 1, 2), ~~~~~
 \psi_{HR} = -\tau^2 q_{HR} = -\tau^2 \left( \begin{array}{c} u_{HR} \\ d_{HR}
 \end{array} \right),
\end{eqnarray}
which transform under $T$ parity as $\Psi_1 \longrightarrow -\Sigma_0 \Psi_2$,
$\Psi_2 \longrightarrow -\Sigma_0 \Psi_1$ and $\Psi_{HR} \longrightarrow -\Psi_{HR}$,
where $\tau^2$ is the second Pauli matrix. $q_i~(i=1,2)$ are the doublets under
$SU(2)_i$, and $T$ parity exchanges $q_1$ and $q_2$.

\par
The transformations for $\Psi_1$, $\Psi_2$ and $\Psi_{HR}$ under
$SU(5)$ are as $\Psi_1 \longrightarrow V^{*} \Psi_1$, $\Psi_2
\longrightarrow V \Psi_2$ and $\Psi_{HR} \longrightarrow U
\Psi_{HR}$, where $V \in SU(5)$. The matrix $U$ is a function of
both $V$ and the ``pion'' matrix $\Pi$, defined by using the
transformation of $\Sigma$ as
\begin{eqnarray}
 \Sigma \longrightarrow V \Sigma V^T
 ~~~\Longrightarrow~~~
 \xi \longrightarrow V \xi U^{\dag} = U \xi \Sigma_0 V^T \Sigma_0,
\end{eqnarray}
where $\xi = e^{i \Pi/f}$.

\par
Considering the transformation properties of $\Psi_1$, $\Psi_2$ and
$\Psi_{HR}$ under $T$ parity and $SU(5)$, we may construct the
following three $SU(2)$ doublets with definite chirality and
$T$ parity:
\begin{eqnarray}
& q_{SM} = (q_1 - q_2)/\sqrt{2},~~~~~~& (\mbox{$T$ even, left-handed}), \nonumber \\
& q_{H} = (q_1 + q_2)/\sqrt{2},~~~~~~& (\mbox{$T$ odd, left-handed}), \\
& q_{HR},~~~~~& (\mbox{$T$ odd, right-handed}) \nonumber
\end{eqnarray}
The $T$-even $SU(2)$ doublet $q_{SM}$ is identified with the SM
left-handed quark doublet, while $q_H$ and $q_{HR}$ are left- and
right-handed mirror quark doublets with odd $T$ parity. Via the
Lagrangian
\begin{eqnarray}
 {\cal L}_{mirror} = -\kappa f \Big( \bar{\Psi}_2 \xi +
 \bar{\Psi}_1 \Sigma_0 \Omega \xi^{\dag} \Omega \Big)
 \Psi_{HR} ~+ ~{\rm H.c.},
\end{eqnarray}
and the $T$-odd mirror quark $q_-$, a Dirac fermion doublet defined as
$\left( q_- \right)_L = q_H$ and $\left( q_- \right)_R = q_{HR}$,
acquires a mass of $\sqrt{2} \kappa f$ before EWSB. After EWSB, a
small mass splitting between the $T$-odd up- and down-type mirror
quarks is induced, and the masses are given by
\cite{Hubisz:2004ft,Hubisz:2006mass}
\begin{eqnarray}\label{m_Q}
 m_{u_-} \simeq \sqrt{2} \kappa f
 \left(
 1 - \frac{1}{8}\frac{v_{SM}^2}{f^2}
 \right),~~~~
 m_{d_-} = \sqrt{2} \kappa f.
\end{eqnarray}

\par
The $T$-odd mirror quark sector involves two CKM-like unitary mixing
matrices $V_{Hu}$ and $V_{Hd}$, which satisfy
$V_{Hu}^{\dag}V_{Hd}=V_{CKM}$ \cite{Blanke:2007ckm}. The related
couplings of the $T$-odd mirror quarks used in our calculations are
listed in Table \ref{tab1} \cite{Blanke:2007ckm,vertex}. In the
following calculations we take $V_{Hu}$ to be a unit matrix, then we
have $V_{Hd}=V_{CKM}$.
%%%%%%%%%%%%%%%%%%%%%%%%table%%%%%%%%%%%%%%%%%%%%%%%%%%%%%%%%%%%%%%%%
\begin{table}[h]
\begin{center}
\begin{tabular}{|c|l||c|l|}
\hline
Vertex & ~~~~~~~~Feynman rule & Vertex & ~~~~~~~~~Feynman rule \\
\hline
&&& \\
$A_{H}^{\mu} \bar{u}_-^i u^j$ & $-i\left(\frac{g'}{10}+\frac{g}{2}
\sin\theta_H\right)(V_{Hu})_{ij} \gamma^\mu P_L$ &
$A_{H}^{\mu} \bar{d}_-^i d^j$ & $i\left(-\frac{g'}{10}+\frac{g}{2}
\sin\theta_H\right)(V_{Hd})_{ij} \gamma^\mu P_L$\\
&&& \\
$\bar{q}_{-}^{\alpha} q_{-}^{\beta} G^{a}_{\mu}$ & $ig_s (T^a)_{\alpha\beta}\gamma^{\mu}$ && \\
&&& \\
\hline
\end{tabular}
\caption{\label{tab1} The related LHT Feynman rules used in this
work, where $q_-=u_-,d_-,c_-,s_-$, $i$ and $j$ are the generation indices.}
\end{center}
\end{table}

\vskip 5mm
\par
\section{Calculations }\label{calc}
\par
Our main focus will be on the $T$-odd mirror quark production of the
first two generations associated with a heavy photon at the CERN
Large Hadron Collider (LHC) in the framework of the LHT. Based on
the description of the LHT in Sec. II, our analysis only depends
on two free model parameters: the global symmetry-breaking scale $f$
and the flavor-independent Yukawa coupling $\kappa$ in the range of
$0.5 \leq \kappa \leq 1.5$ \cite{Hubisz-1,Hubisz-2}.

\par
In the calculation of the LO cross section and the NLO QCD
corrections, we adopt the 't Hooft-Feynman gauge. The developed
FeynArts 3.4 package \cite{FeynArts} and FormCalc 5.4 program
\cite{FormCalc} are used for Feynman diagram and amplitude
generation and algebraic manipulation. We adopt the four-flavor
scheme (4FS) for the initial parton convolution, and take the $u$,
$d$, $c$ and $s$ quark to be massless ($m_u=m_d=m_c=m_s=0$).

\par
\subsection{LO cross section}
\par
\begin{figure}
\begin{center}
\includegraphics[width=0.44\textwidth]{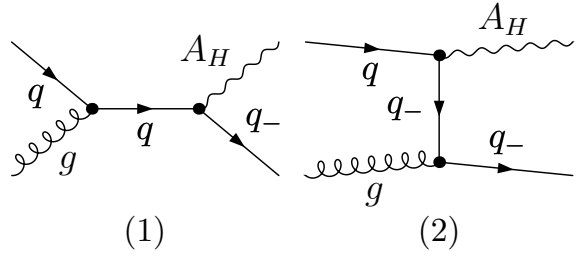}
\caption{ \label{fig1} The tree-level Feynman diagrams for the
partonic process $qg \to A_H q_-$. }
\end{center}
\end{figure}

In the case of no quark mixing between the first two generations and
the third generation (i.e., $V_{ub}=V_{cb}=V_{td}=V_{ts}=0$), only
the following quark-gluon fusion partonic processes contribute to
the $T$-odd mirror quark production associated with a heavy photon
at the LHC,
\begin{eqnarray}
\label{subprocess} ~~q g \to A_H q^{\prime}_-,~~~~~~~~~~ (q
q^{\prime}_- = u u_-, c c_-, \bar{u} \bar{u}_-, \bar{c} \bar{c}_-, d
d_-, d s_-, s d_-, s s_-, \bar{d} \bar{d}_-, \bar{d} \bar{s}_-,
\bar{s} \bar{d}_-, \bar{s} \bar{s}_-).
\end{eqnarray}
Since we take $V_{Hu} = I$ and $V_{Hd} = V_{CKM}$, all the up-type
quark-gluon fusion subprocesses are flavor conserved, and the flavor
changing occurs only in the down-type quark-gluon fusion
subprocesses as shown in Eq.(\ref{subprocess}). For each down-type
quark-gluon fusion subprocess, the amplitude squared is proportional
to the CKM matrix element squared induced by the
$A_H$-$q_-$-$\bar{q}$ coupling. However, due to the unitarity of the
CKM matrix, the sum of the production rates for all the subprocesses
with the same initial parton flavor is free of the CKM matrix
element. For example, the cross sections for the partonic processes
$d g \to A_H d_-$ and $d g \to A_H s_-$ are proportional to
$|V_{ud}|^2$ and $|V_{cd}|^2$, respectively, but the summation of
the cross sections for these two subprocesses is independent of the
CKM matrix element due to the fact that $|V_{ud}|^2 + |V_{cd}|^2 =
1$. That is to say, we may consider only the flavor-conserved
subprocesses by taking $V_{CKM} = I$ and thus get the right results
in the calculation of the total cross section. In the following LO
and NLO calculations, we adopt this strategy by taking $V_{CKM} =
I$, and denote the partonic processes contributing to the $A_Hq_-$
associated production at the LHC as
\begin{eqnarray}
q(p_{1}) + g(p_{2}) \to A_H(p_{3}) + q_-(p_{4}),
\end{eqnarray}
where $q q_- = u u_-, \bar{u} \bar{u}_-, d d_-, \bar{d} \bar{d}_-, c
c_-,\bar{c} \bar{c}_-, s s_-, \bar{s} \bar{s}_-$, and $p_i~(i = 1,
2, 3, 4)$ represent the four-momenta of the incoming and the
outgoing particles, respectively. The tree-level Feynman diagrams
for the partonic process $q g \to A_H q_-$ are presented in
Fig.\ref{fig1}.

\par
The LO cross section for the partonic process $q g \to A_H q_-$ can
be expressed as
\begin{eqnarray}
\hat{\sigma}_{LO}^{q g}= \frac{1}{4}\frac{1}{24}\frac{(2 \pi )^4}{4|\vec{p}|\sqrt{\hat{s}}}\int
\sum_{spin}\sum_{color} |{\cal M}_{LO}^{q g}|^2 d\Omega_2,&& (q = u
,\bar u,d,\bar d, c ,\bar c, s,\bar s),
\end{eqnarray}
where $\vec{p}$ is the three-momentum of one initial parton in the
center-of-mass system (c.m.s.), the factors $\frac{1}{4}$ and
$\frac{1}{24}$ come from the averaging over the spins and colors of
initial partons respectively, $\sqrt{\hat{s}}$ is the partonic
c.m.s. colliding energy, and ${\cal M}_{LO}^{q g}$ is the LO
amplitude for the partonic process $q g \to A_H q_-$. The two
summations are taken over the spins and colors of all the relevant
initial and final particles, separately. The integration is
performed over the two-body phase space of the final particles $A_H$
and $q_-$, where the phase space element $d\Omega_2$ is defined as
\begin{eqnarray}
d\Omega_2 = \delta^{(4)} (p_1 + p_2 - p_3 - p_4) \frac{d^3
\vec{p}_3}{(2 \pi)^3 2 E_3} \frac{d^3 \vec{p}_4}{(2 \pi)^3 2 E_4}.
\end{eqnarray}
Then the LO total cross section for the $p p \to A_H q_- + X~ (q_- =
u_-, \bar{u}_-, d_-, \bar{d}_-, c_-, \bar{c}_-, s_-, \bar{s}_-)$
process can be obtained as
\begin{eqnarray}
\label{pp-total cross section}
&& \sigma_{LO}(pp \to A_H q_- + X) \nonumber \\
&& = \sum_{q=u,d,c,s,}^{\bar{u},\bar{d},\bar{c},\bar{s}} \left\{
\int dx_A dx_B \Big[ G_{q/A}(x_A, \mu_f) G_{g/B}(x_B, \mu_f)
\hat{\sigma}_{LO}^{q g}(q g \to A_H q_-, x_{A} x_{B} s, \mu_f) + (A
\leftrightarrow B) \Big] \right\},~~~~~~~
\end{eqnarray}
where $G_{i/P}~ (i=q,g,~P=A,B)$ represents the PDF of parton $i$ in
proton $P$, $x_P$ is the momentum fraction of a parton (quark or
gluon) in proton $P$, and $\mu_f$ and $\mu_r$ are the factorization
and renormalization scales, respectively.
%%%%%%%%%%%%%%%%%%%%%%%%%%%%%figure1%%%%%%%%%%%%%%%%%%%%%%%%%%%%%%%%%

\par
\subsection{ NLO QCD corrections  }
\par
It is known that the NLO QCD corrections to any hadronic
process include three components: (1) loop virtual correction, (2)
real gluon/light-quark emission correction, and (3) PDF
counterterms. To regularize the ultraviolet (UV) and infrared (IR)
divergences, we adopt the dimensional regularization scheme in $D =
4 - 2 \epsilon$ dimensions.

\par
{\bf 1. Virtual correction }
\par
\begin{figure}
\begin{center}
\includegraphics[width=0.66\textwidth]{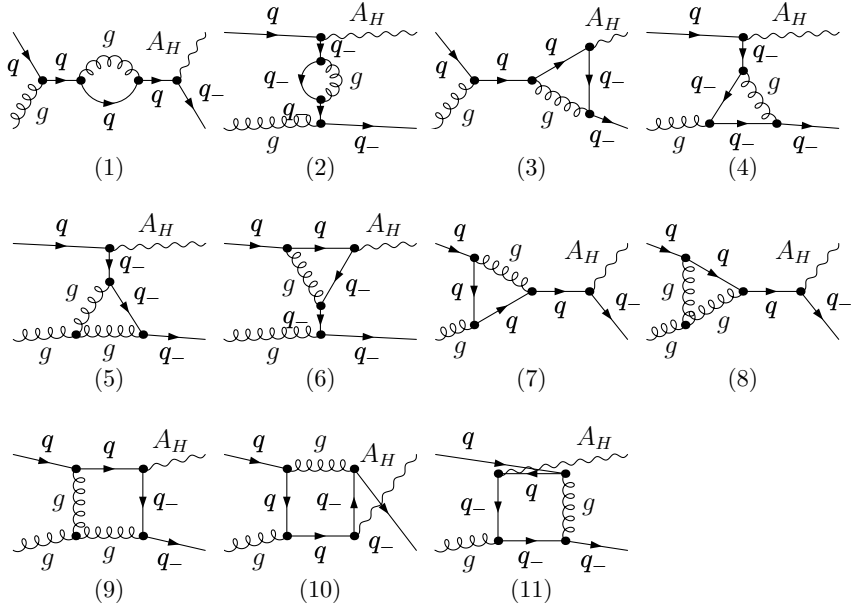}
\caption{ \label{fig2} The QCD one-loop Feynman diagrams for the
partonic process $q g \to A_H q_-$. }
\end{center}
\end{figure}

We present the QCD one-loop Feynman diagrams for the partonic
process $q g \to A_H q_-$ in Fig.\ref{fig2}. In the calculations of
the QCD one-loop virtual correction, we will meet both UV and IR
singularities. To remove the UV divergences, the strong coupling
constant, the wave functions and masses of related colored particles
should be renormalized by introducing the renormalization constants
$\delta g_s$, $\delta Z_g$, $\delta Z_q^{L,R}$, $\delta
Z_{q_-}^{L,R}$ and $\delta m_{q_-}$. These
renormalization constants are defined as
\begin{eqnarray}
& G_{\mu}^0 = \Big(1 + \frac{1}{2} \delta Z_g \Big) G_{\mu}~, &~~~~~~~ g_s^0 = g_s + \delta g_s~, \nonumber \\
&~ \psi^{0,L,R}_{q} = \Big(1 + \frac{1}{2}
\delta Z_{q}^{L,R} \Big) \psi^{L,R}_{q}~, & \nonumber \\
& ~ \psi^{0,L,R}_{q_-} = \Big(1 + \frac{1}{2} \delta Z_{q_-}^{L,R} \Big)\psi^{L,R}_{q_-}, &
~~~ m^0_{q_-} = m_{q_-} + \delta m_{q_-},
\end{eqnarray}
where we denote bare fields and constants by an index 0,
$g_s$ is the strong coupling constant, $G_{\mu}$,
$\psi^{L,R}_{q}$ and $\psi^{L,R}_{q_-}$ represent the fields of
gluon, quark and $T$-odd mirror quark, respectively, and $m_{q_-}$
denotes the mass of $T$-odd mirror quark. We adopt the on-shell
renormalization scheme to fix the mass and wave function
renormalization constants, and then obtain
\begin{eqnarray}
&&\delta Z_g = - \frac{\alpha_s(\mu_r)}{2 \pi}
             \left[
             \frac{3}{2} \Delta_{UV} + \frac{5}{6} \Delta_{IR} + \frac{1}{3} \ln \frac{\mu_r^2}{m_t^2}
             + \frac{1}{3} \sum_{T=T_+}^{T_-} \ln \frac{\mu_r^2}{m_{T}^2}
             + \frac{1}{3} \sum_{q = u, c, t}^{d, s, b} \ln \frac{\mu_{r}^2}{m_{q_-}^2} \right], \nonumber \\
&&\delta Z_{q}^{L,R} = - \frac{\alpha_s(\mu_r)}{3 \pi}\Big[ \Delta_{UV} - \Delta_{IR} \Big], \nonumber \\
&&\delta Z_{q_-}^{L,R} = - \frac{\alpha_s(\mu_r)}{3 \pi}
                     \left[
                     \Delta_{UV} + 2\Delta_{IR} + 4 + 3 \ln \frac{\mu_r^2}{m_{q_-}^2}
                     \right], \nonumber \\
&&\frac{\delta m_{q_-}}{m_{q_-}} = - \frac{\alpha_s(\mu_r)}{3 \pi}
                                 \left[
                                 3 \left( \Delta_{UV} + \ln \frac{\mu_r^2}{m_{q_-}^2} \right) + 4
                                 \right],
\end{eqnarray}
where $\Delta_{UV} = \frac{1}{\epsilon_{UV}} - \gamma_E + \ln
(4\pi)$, $\Delta_{IR} = \frac{1}{\epsilon_{IR}} - \gamma_E + \ln
(4\pi)$, $T_+$ is the vectorlike top quark, and $T_-$ is the
$T$-partner of $T_+$. For the renormalization of the strong coupling
constant, we adopt the $\overline{MS}$ scheme at the renormalization
scale $\mu_{r}$, except that the divergences associated with the
massive top quark, $T$-odd mirror quarks ($u_-, d_-, c_-, s_-, t_-,
b_-$) and $T_{\pm}$ loops are subtracted at zero momentum \cite{gs}.
Then the renormalization constant $\delta g_s$ can be obtained as
\begin{eqnarray}
\frac{\delta g_s}{g_s} = -\frac{\alpha_s(\mu_r)}{4\pi}
                         \left[
                         \frac{3}{2} \Delta_{UV} + \frac{1}{3}\ln\frac{m_{t}^2}{\mu_{r}^2}
                         + \frac{1}{3} \sum_{T=T_+}^{T_-} \ln \frac{m_{T}^2}{\mu_{r}^2}
                         + \frac{1}{3} \sum_{q = u, c, t}^{d, s, b} \ln \frac{m_{q_-}^2}{\mu_{r}^2}
                         \right].
\end{eqnarray}
It is obvious that the terms of $\ln\frac{m_{x}^2} {\mu_{r}^2}$
$(x = T_+, T_-, t_-, b_-)$ contributed by the
renormalization constants $\delta Z_g$ and $\delta g_s/g_s$ exactly
cancel each other in the QCD NLO counterterm amplitude.
Therefore, the NLO QCD correction is independent of $m_{T_{+}}$,
$m_{T_{-}}$, $m_{t_-}$ and $m_{b_-}$.

\par
After performing the renormalization procedure, the virtual
correction is UV finite. However, it still contains soft and
collinear IR divergences, which can be eliminated by including the
contributions of the real gluon/light-quark emission subprocesses
and the PDF counterterms.
%%%%%%%%%%%%%%%%%%%%%%%%%%%figure2%%%%%%%%%%%%%%%%%%%%%%%%%%%%%%%%%%

\par
{\bf 2. Real emission correction }
\par
\begin{figure}
\begin{center}
\includegraphics[width=0.7\textwidth]{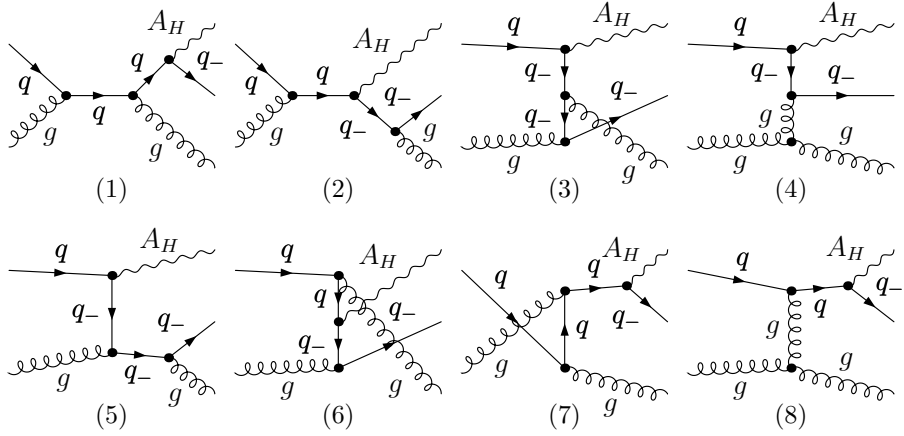}
\caption{ \label{fig3} The tree-level Feynman diagrams for the real
gluon emission partonic process $q g \to A_H q_- + g$. }
\end{center}
\end{figure}

We denote the real gluon emission partonic processes for the $A_H
q_-$ associated production as
\begin{eqnarray}
q(p_{1}) + g(p_{2}) \to A_H(p_{3}) + q_-(p_{4}) + g(p_{5}), ~~~~~~~
(q = u, d, c, s, \bar{u}, \bar{d}, \bar{c}, \bar{s}),
\end{eqnarray}
and plot the related Feynman diagrams in Fig.\ref{fig3}. We employ
the two cutoff phase space slicing (TCPSS) method \cite{19} to
isolate soft and collinear IR singularities of these partonic
processes. In adopting this method, two arbitrary cutoffs, $\delta_s$
and $\delta_c$, are introduced to separate the phase space of $q g
\to A_H q_- + g$ into soft ($E_5 \leq \frac{1}{2} \delta_s
\sqrt{\hat{s}}$), hard collinear ($E_5 > \frac{1}{2}\delta_s
\sqrt{\hat{s}}$, $\hat{s}_{15}~ \mbox{or}~ \hat{s}_{25} \leq
\delta_c \hat{s}$) and hard noncollinear ($E_5 > \frac{1}{2}\delta_s
\sqrt{\hat{s}}$, $\hat{s}_{15}~ \mbox{and}~ \hat{s}_{25} > \delta_c
\hat{s}$) regions, where $\hat{s}_{ij}=(p_i+p_j)^2$. Then the cross
section for the real gluon emission partonic process $q g \to A_H
q_- + g$ is expressed as
\begin{eqnarray}
\hat{\sigma}_g = \hat{\sigma}^S_g + \hat{\sigma}^{HC}_g +
\hat{\sigma}^{\overline{HC}}_g,
\end{eqnarray}
where the superscripts $S$, $HC$ and $\overline{HC}$ stand for the
soft, hard collinear and hard noncollinear regions, respectively.
The soft correction $\hat{\sigma}_g^S$ and the hard collinear
correction $\hat{\sigma}^{HC}_g$ contain soft and collinear IR
singularities, respectively, while the hard noncollinear correction
$\hat{\sigma}^{\overline{HC}}_g$ is IR finite. According to the
Kinoshita-Lee-Nauenberg (KLN) \cite{KLN} theorem, the soft IR
singularity in $\hat{\sigma}^{S}_g$ can be canceled exactly by that
in the virtual correction. The collinear IR singularity in
$\hat{\sigma}^{HC}_g$ can be partially canceled by that in the
virtual correction, and the remained collinear IR divergence will be
absorbed by the PDF counterterms.
%%%%%%%%%%%%%%%%%%%%%%%%%%figure3%%%%%%%%%%%%%%%%%%%%%%%%%%%%%%%%%%%

\par
All the real light-quark emission partonic processes for the $A_H
q_-$ associated production are listed below:
\begin{eqnarray}
&& q(p_1) + q^{\prime}(p_2) \to A_H(p_3) + q_-(p_4) +
q^{\prime}(p_5), ~~~~~(q'\neq \bar{q}),  \nb \\
&& q'(p_1)+\bar{q}'(p_2) \to A_H(p_3)+q_{-}(p_4)+\bar{q}(p_5), \nb \\
&& g(p_{1})+ g(p_{2})\to A_H(p_{3})+q_-(p_{4})+\bar{q}(p_{5}),
\end{eqnarray}
where $q, q^{\prime} = u, d, c, s, \bar{u}, \bar{d}, \bar{c},
\bar{s}$. Using the TCPSS method, the phase space of a real
light-quark emission partonic process is decomposed into collinear
($\hat{s}_{15}~ \mbox{or}~ \hat{s}_{25} \leq \delta_c \hat{s}$) and
noncollinear ($\hat{s}_{15}~ \mbox{and}~ \hat{s}_{25} > \delta_c
\hat{s}$) regions, and then the cross section is expressed as
\begin{eqnarray}
\hat{\sigma}_{q} = \hat{\sigma}_{q}^{C} + \hat{\sigma}_{q}^{\overline{C}}.
\end{eqnarray}
The noncollinear correction $\hat{\sigma}_{q}^{\overline{C}}$ is IR finite,
while the collinear correction $\hat{\sigma}_{q}^{C}$ contains collinear IR
singularity which can be canceled by the corresponding PDF counterterms.
%%%%%%%%%%%%%%%%%%%%%%%%%%%%%%%%%%%%%%%figure4%%%%%%%%%%%%%%%%%%%%%
\begin{figure}
\begin{center}
\includegraphics[width=0.7\textwidth]{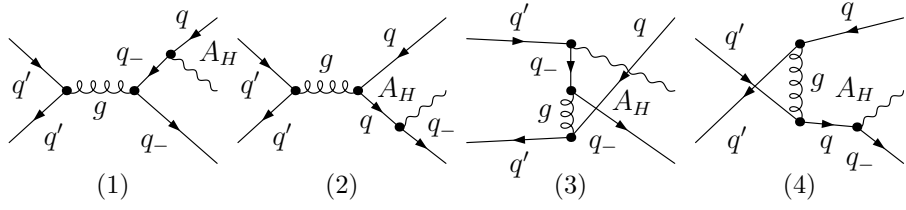}
\caption{ \label{fig4} The tree-level Feynman diagrams for the real
light-quark emission partonic process $q^{\prime} \bar{q}^{\prime}
\to A_H q_- + \bar{q}$. Figures.\ref{fig4}(3)-\ref{fig4}(4) only appear in the case
of $q^{\prime} = q$.}
\end{center}
\end{figure}
%%%%%%%%%%%%%%%%%%%%%%%%%%%%%%%%%%%%figure5%%%%%%%%%%%%%%%%%%%%%%%%
\begin{figure}
\begin{center}
\includegraphics[width=0.7\textwidth]{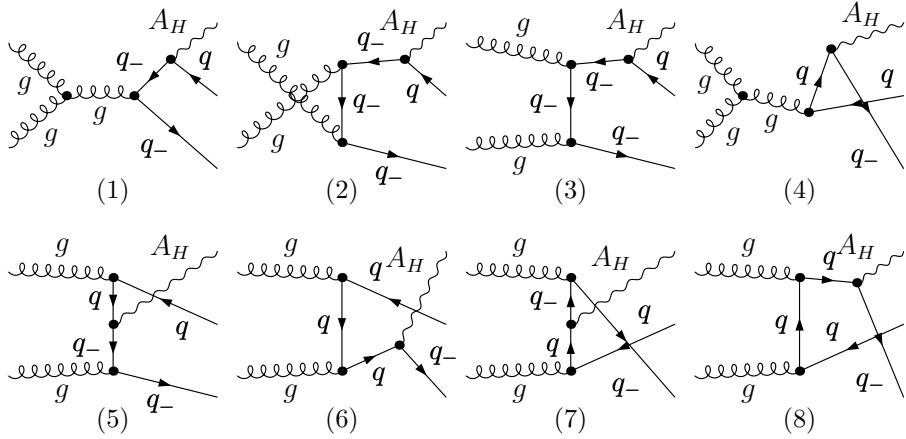}
\caption{\label{fig5} The tree-level Feynman diagrams for the real
light-quark emission partonic process $g g \to A_H q_- +\bar{q}$. }
\end{center}
\end{figure}

\par
Among all the real light-quark emission partonic processes, only
$q^{\prime} \bar{q}^{\prime} \to A_H q_- + \bar{q}$ and $g g \to A_H
q_- + \bar{q}$ may have resonance effect. We present the tree-level
Feynman diagrams for these partonic processes in Figs.\ref{fig4} and
\ref{fig5}, respectively, and find that Fig.\ref{fig4}(1) and
Figs.\ref{fig5}(1)-\ref{fig5}(3)contain on-shell $q_-$ contributions. To deal
with the $q_-$ resonance effect, we replace $m_{q_-}^2$ by
$m_{q_-}^2 - i m_{q_-} \Gamma_{q_-}$ for all possible on-shell $q_-$
propagators in Fig.\ref{fig4}(1) and Figs.\ref{fig5}(1)-\ref{fig5}(3). However,
this resonance effect will induce extremely large correction and
eventually destroy the perturbative convergence. Furthermore,
Fig.\ref{fig4}(1) and Figs.\ref{fig5}(1)-\ref{fig5}(3) are also counted towards
the $T$-odd mirror quark pair production partonic processes
$q^{\prime} \bar{q}^{\prime} \to q_- \bar{q}_-$ and $g g \to q_-
\bar{q}_-$, respectively, followed by an on-shell decay $\bar{q}_-
\to \bar{q}A_H$. Therefore, we adopt the PROSPINO subtraction
strategy \cite{PROSPINO-ref,on-shell subtraction} for the
$q^{\prime} \bar{q}^{\prime} \to A_H q_- + \bar{q}$ and $g g \to A_H
q_- + \bar{q}$ partonic processes to avoid double counting and keep
the convergence of perturbative calculations. The PROSPINO
subtraction scheme is performed by making a replacement of the
Breit-Wigner propagator:
\begin{eqnarray}
 \frac{|{\cal M}|^2( s_{q_-} )}{( s_{q_-} - m_{q_-}^2 )^2
 + m_{q_-}^2 \Gamma_{q_-}^2}
 & \to &
 \frac{|{\cal M}|^2( s_{q_-} )}{( s_{q_-} - m_{q_-}^2 )^2
 + m_{q_-}^2 \Gamma_{q_-}^2} \nonumber \\
 &&-
 \frac{|{\cal M}|^2( m_{q_-}^2 )}{( s_{q_-} - m_{q_-}^2 )^2
 + m_{q_-}^2 \Gamma_{q_-}^2}
 \Theta( \hat{s} - 4 m_{q_-}^2 )
 \Theta( m_{q_-} - m_{A_H} ),
\end{eqnarray}
where $s_{q_-}$ is the squared momentum flowing through the
intermediate $q_-$ propagator.

\par
After adding the renormalized virtual correction with the
contributions of the real gluon/light-quark emission processes and
the PDF counterterms, $\delta G_{i/P}(x,\mu_f)$ ($i=g,u,\bar
u,d,\bar d,c,\bar c, s,\bar s$), together \cite{vertex,19}, the UV
and IR singularities are exactly vanished. These cancelations are
verified both analytically and numerically in our calculations.

\par
{\bf 3. Total NLO QCD correction}
\par
The total NLO QCD corrected cross section for the hadronic $T$-odd
mirror quark production associated with a heavy photon can be
expressed as
\begin{eqnarray}
\label{TotalCorr} \sigma_{NLO} = \sigma_{LO} + \Delta \sigma_{NLO} =
\sigma_{LO} + \Delta \sigma^{(2)} + \Delta\sigma^{(3)}.
\end{eqnarray}
The two-body QCD correction $\Delta \sigma^{(2)}$ includes the
one-loop virtual correction, the cross sections for the real gluon
emission processes over the soft and hard collinear phase space
regions and the cross sections for the real light-quark emission
processes over the collinear phase space regions,  while the
three-body QCD correction $\Delta \sigma^{(3)}$ contains the cross
sections over the hard noncollinear regions for the real gluon
emission processes and the noncollinear regions for the real
light-quark emission processes.

\vskip 5mm
\section{Numerical results and discussions }\label{numres}
\par
\subsection{Input parameters}
\par
In our numerical calculations we set $V_{Hu}  = I$, $V_{Hd}= V_{CKM}
= I$ and treat light quarks as massless particles. The SM
electroweak input parameters are taken as $\alpha_{{\rm ew}}^{-1} =
137.036$, $m_W = 80.385~{\rm GeV}$, $m_Z = 91.1876~{\rm GeV}$ and
$\sin^2\theta_W = 1-\left(\frac{m_W}{m_Z}\right)^2 = 0.2229$
\cite{databook}. The c.m.s. energies of proton-proton collision for
the future and present LHC are taken as $\sqrt{s} = 14~{\rm TeV}$
and $\sqrt{s} = 8~{\rm TeV}$. We adopt CTEQ6L1 and CTEQ6M PDFs for
the initial state convolution in the LO and NLO calculations,
respectively \cite{cteq}. $\alpha_s(\mu)$ is determined by the QCD
parameter $\Lambda_5^{LO} = 165~{\rm MeV}$ for the CTEQ6L1 at the LO
and $\Lambda_5^{\overline{MS}} = 226~{\rm MeV}$ for the CTEQ6M at
the NLO \cite{databook}. The factorization and renormalization
scales are set to be equal ($ \mu = \mu_f = \mu_r$) and the central
value is taken as $\mu_0 = (m_{A_H} + m_{d_-})/2$. In Table
\ref{tab2} we list the masses of $A_H$ and $T$-odd mirror quarks for
some typical values of the global symmetry-breaking scale $f$ with
$\kappa = 0.5$, $1.0$, and $1.5$, separately.
%%%%%%%%%%%%%%%Table%%%%%%%%%%%%%%%%%%%%%%%%%%%%%%%%%%%%%%%%%%%%%%%%%%
\begin{table}
\begin{center}
\begin{tabular}{|c|c|c|c|c|}
  \hline
      $\kappa$ &  $f~({\rm GeV})$  &  $m_{A_H}~({\rm GeV})$  & $m_{u_-}=m_{c_-}~ ({\rm GeV})$
      & $m_{d_-}=m_{s_-}~ ({\rm GeV})$  \\
  \hline
         &  600  & 82.5  & 415.3  & 424.3  \\
    0.5  &  900  & 131.8 & 630.5  & 636.4  \\
         &  1200 & 179.5 & 844.1  & 848.5  \\
         &  1500 & 226.6 & 1057.1 & 1060.7 \\
  \hline
         &  600  & 82.5  & 830.7  & 848.5   \\
    1.0  &  900  & 131.8 & 1260.9 & 1272.8  \\
         &  1200 & 179.5 & 1688.1 & 1697.1  \\
         &  1500 & 226.6 & 2114.2 & 2121.3  \\
  \hline
         &  600  & 82.5  & 1246.0 & 1272.8  \\
    1.5  &  900  & 131.8 & 1891.4 & 1909.2  \\
         &  1200 & 179.5 & 2532.2 & 2545.6  \\
         &  1500 & 226.6 & 3171.3 & 3182.0  \\
  \hline
\end{tabular}
\end{center}
\begin{center}
\begin{minipage}{15cm}
\caption{\label{tab2} The masses of $A_H$ and $q_-~
(q_- = u_-, d_-, c_-, s_-)$ for some typical values of the global symmetry-breaking scale
$f$ with $\kappa = 0.5, 1.0$ and $1.5$. }
\end{minipage}
\end{center}
\end{table}
%%%%%%%%%%%%%%%%%%%%%%%%%%%%%%%%%%%%%%%%%%%%%%%%%%%%%%%%%%%%%%%%

\par
\subsection{Checks}
\par
The correctness of our calculations is verified in the following aspects:
\par
{\bf 1.} We adopt the same PDFs and input parameters as used in
Ref.\cite{cpyuan:2006ph} and find that our LO cross sections are
in good agreement with the results given in Fig.8 of
Ref.\cite{cpyuan:2006ph}.

\par
{\bf 2.} After summing up all the contributions at the QCD NLO, the
cancelations of UV and IR divergences are verified both analytically
and numerically.

\par
{\bf 3.} We perform the verification of the $\delta_s/\delta_c$
independence of the total NLO QCD correction. The numerical results
show that the total NLO QCD correction $\Delta \sigma_{NLO}$ is
independent of the two cutoffs within the statistical errors. This
independence is an indirect check for the correctness of our work.
In further numerical calculations, we fix $\delta_s = 1 \times
10^{-4}$ and $\delta_c = 1 \times 10^{-6}$.

\par
\subsection{Dependence on factorization and renormalization scale $\mu$ }
\par
\begin{figure}
\begin{center}
\includegraphics[width=0.48\textwidth]{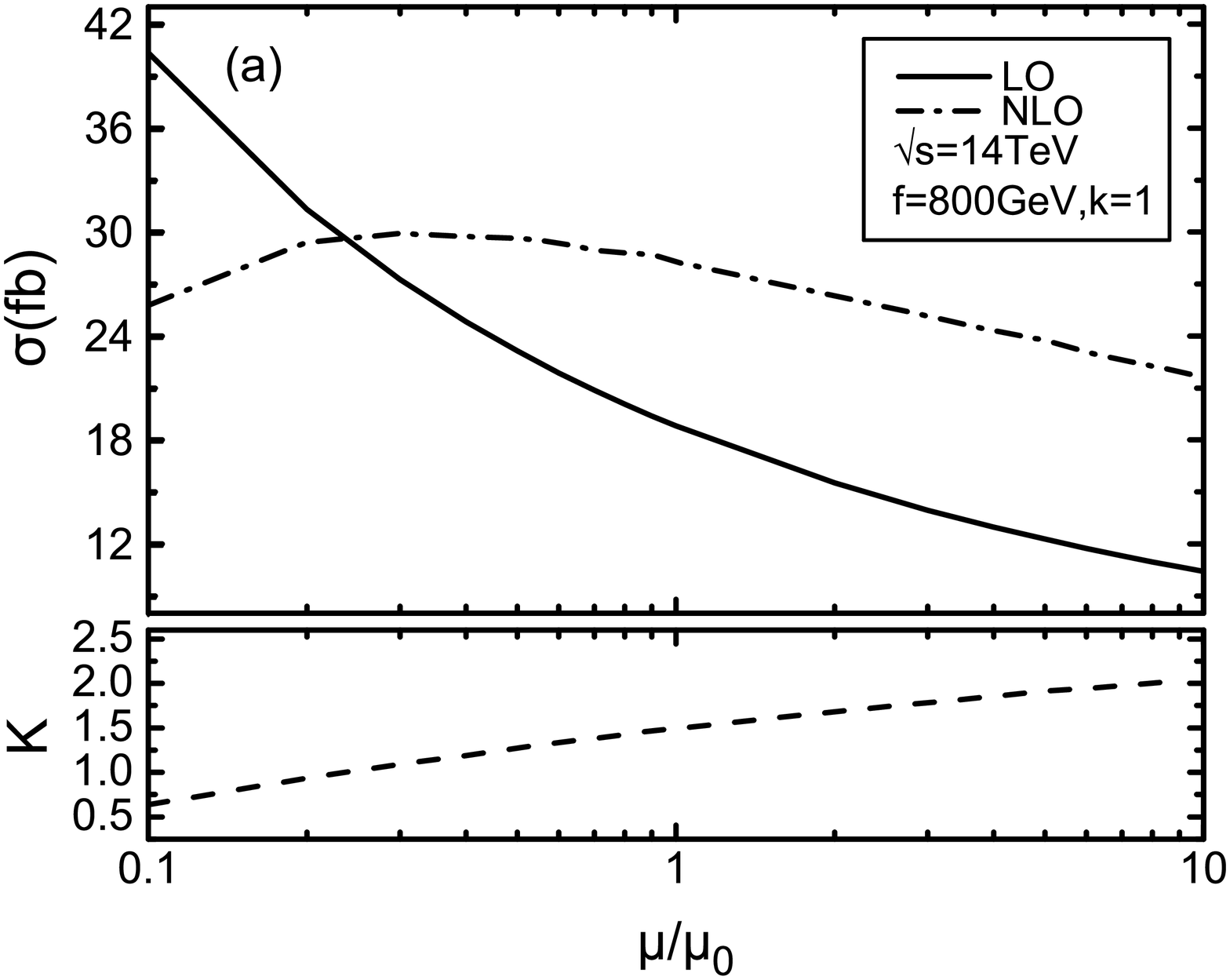}
\includegraphics[width=0.48\textwidth]{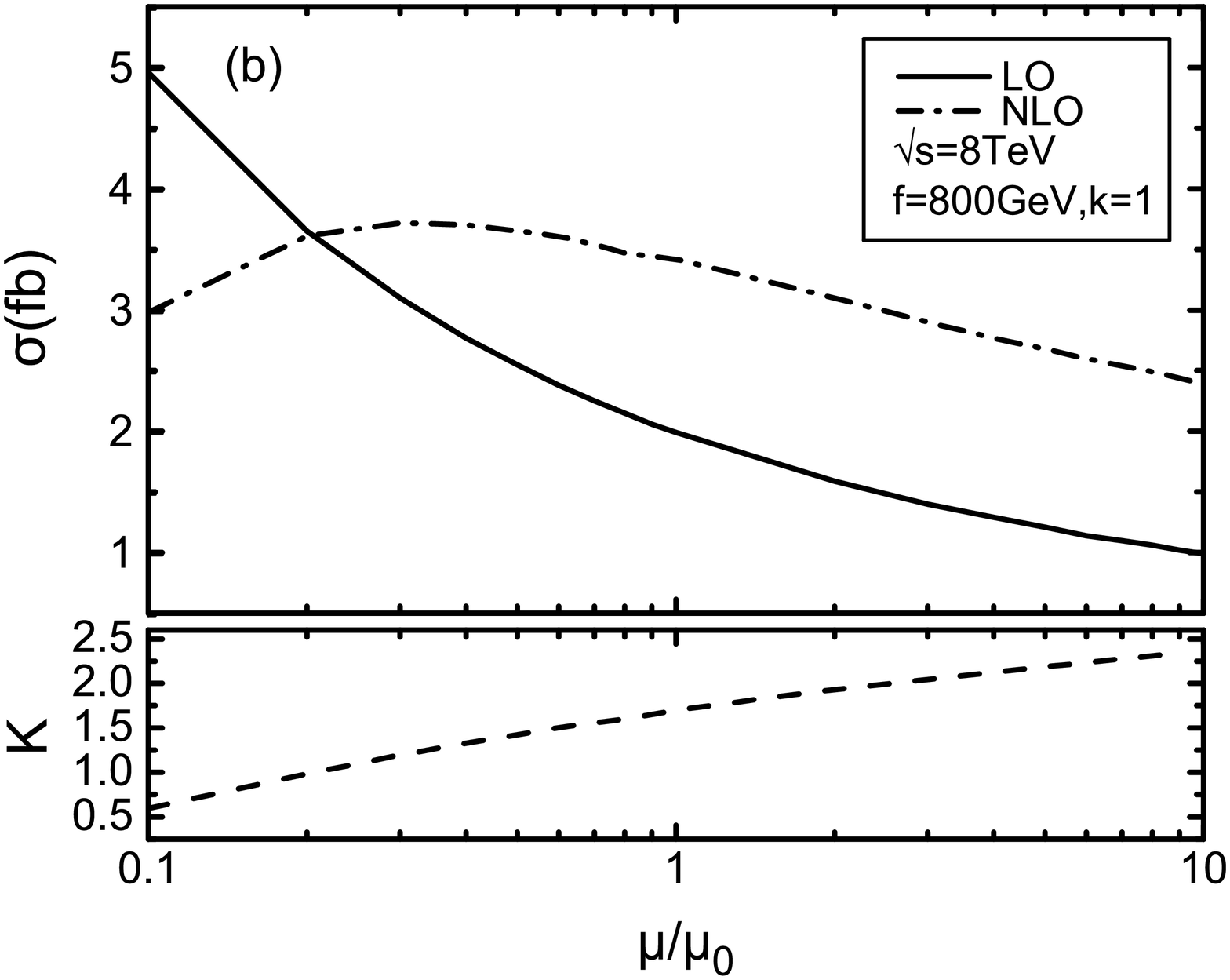}
\caption{ \label{fig6} The dependence of the LO, NLO QCD corrected
integrated cross sections and the corresponding $K$ factors on the
factorization and renormalization scale for the $p p \to A_H q_- + X$
process at the LHC. (a) $\sqrt{s} = 14~{\rm TeV}$. (b) $\sqrt{s} =
8~{\rm TeV}$. }
\end{center}
\end{figure}

In Figs.\ref{fig6}(a) and \ref{fig6}(b) we present the dependence
of the LO, QCD NLO corrected integrated cross sections and the
corresponding $K$ factors on the factorization and renormalization scale
$\mu (\equiv \mu_f = \mu_r)$ for the $p p \to A_H q_- + X~ (q_- =
u_-, d_-, c_-, s_-, \bar{u}_-, \bar{d}_-, \bar{c}_-, \bar{s}_-)$
process at the $\sqrt{s} = 14~{\rm TeV}$ and $8~{\rm TeV}$ LHC,
respectively, by taking $f = 800~ {\rm GeV}$ and $\kappa = 1$. The
numerical results of the cross sections and the corresponding
$K$-factors for some typical values of $\mu$ are listed in Table
\ref{tab3}. From Fig.\ref{fig6}(a) and \ref{fig6}(b) we can find
that the NLO QCD correction reduces the
factorization and renormalization scale dependence of the LO cross
section significantly. At the $\sqrt{s}=14~{\rm TeV}$ and $8~{\rm
TeV}$ LHC, the relative NLO QCD corrections at the central scale
$\mu_0$ are about $49\%$ and $70\%$, respectively. In the following
calculations, we fix the factorization and renormalization scale as
$\mu=\mu_0$.
%%%%%%figure7%%%%%%%%%%%%%%%%%%%%%%%%%%%%%%%%%%%%%%%%%%%%%%%%%
%%%%%%%%%%%%%%%Table%%%%%%%%%%%%%%%%%%%%%%%%%%%%%%tabel3%%%%%%%%%%%%%%%%%%%%%%%%5%5
\begin{table}
\begin{center}
\begin{tabular}{|c|c|c|c|c|}
\hline
$\sqrt{s}~({\rm TeV})$ &  ~$\mu/\mu_0$~  &  ~~$\sigma_{LO}~ (fb)$~~ &  ~$\sigma_{NLO}~ (fb)$~ & ~~~$K$~~~  \\
\hline
               &   0.1          &    40.501(3)    &  25.7(9)        & 0.64  \\
               &   0.5          &    23.281(2)    &  29.6(6)        & 1.27  \\
       14      &    1           &    18.946(1)    &  28.3(6)        & 1.49  \\
               &    5           &    12.4139(9)   &  23.7(4)        & 1.91  \\
               &   10           &    10.5520(8)   &  21.6(4)        & 2.04  \\
\hline
               &   0.1          &    4.9801(4)    &  2.97(5)        & 0.60  \\
               &   0.5          &    2.5653(2)    &  3.65(2)        & 1.42  \\
       8       &    1           &    2.0106(1)    &  3.41(2)        & 1.70  \\
               &    5           &    1.2252(1)    &  2.60(1)        & 2.23  \\
               &   10           &    1.01426(8)   &  2.40(1)        & 2.36  \\
\hline
\end{tabular}
\end{center}
\begin{center}
\begin{minipage}{15cm}
\caption{\label{tab3}  The numerical results of $\sigma_{LO}$,
$\sigma_{NLO}$ and the corresponding $K$ factors for the $p p \to
A_H q_- + X$ process at the $\sqrt{s}=14~{\rm TeV}$ and $8~{\rm
TeV}$ LHC for some typical values of $\mu$, where $f = 800~ {\rm
GeV}$ and $\kappa = 1$.}
\end{minipage}
\end{center}
\end{table}
%%%%%%%%%%%%%%%%%%%%%%%%%%%%%%%%%%%%%%%%%%%%%%%%%%%%%%%%%%%%%%%%%%%%%%%%%%%%%%%%%%%%%%%%%%

\par
\subsection{Dependence on global symmetry-breaking scale $f$ }
\par
\begin{figure}
\begin{center}
\includegraphics[width=0.48\textwidth]{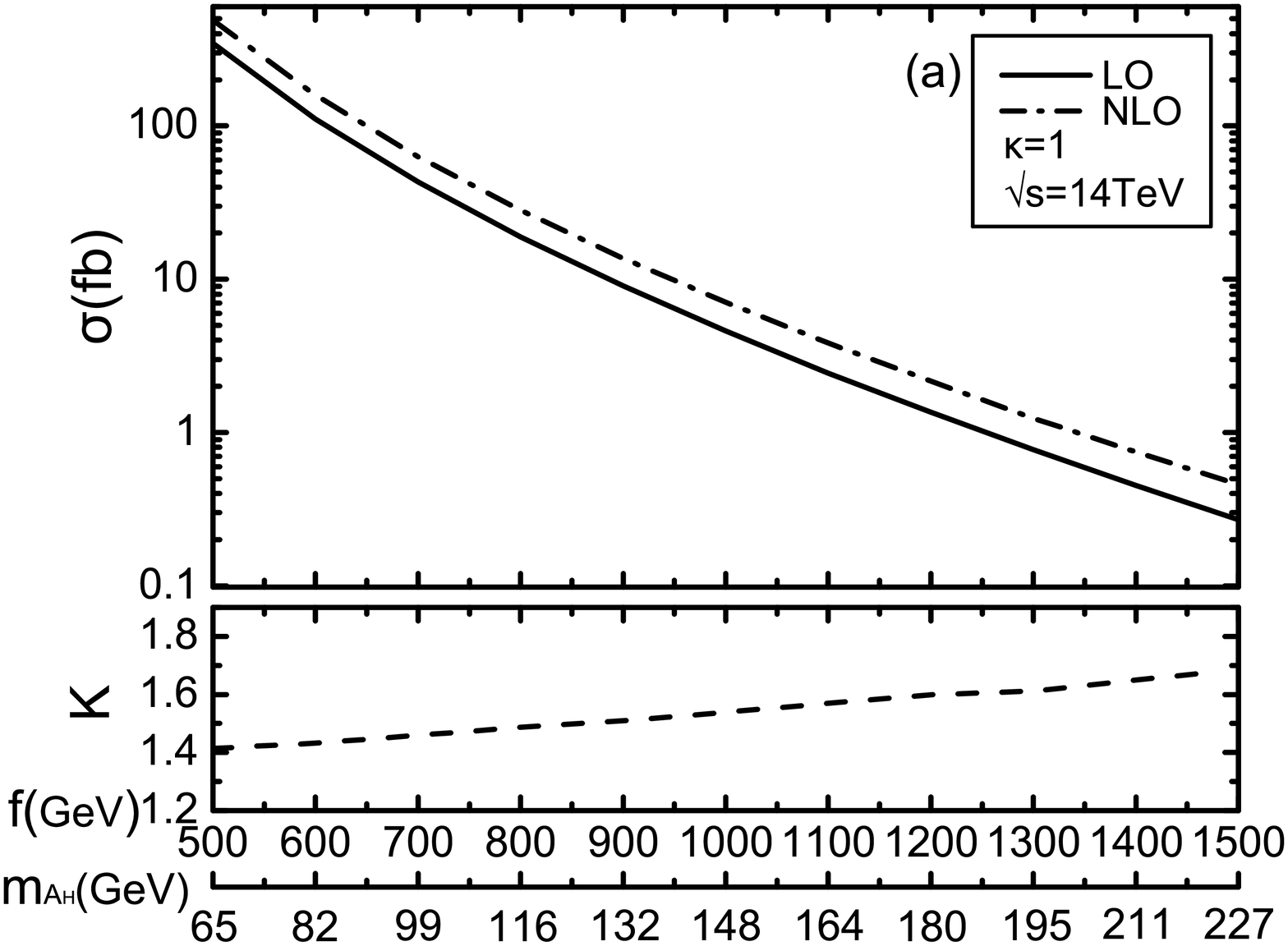}
\includegraphics[width=0.48\textwidth]{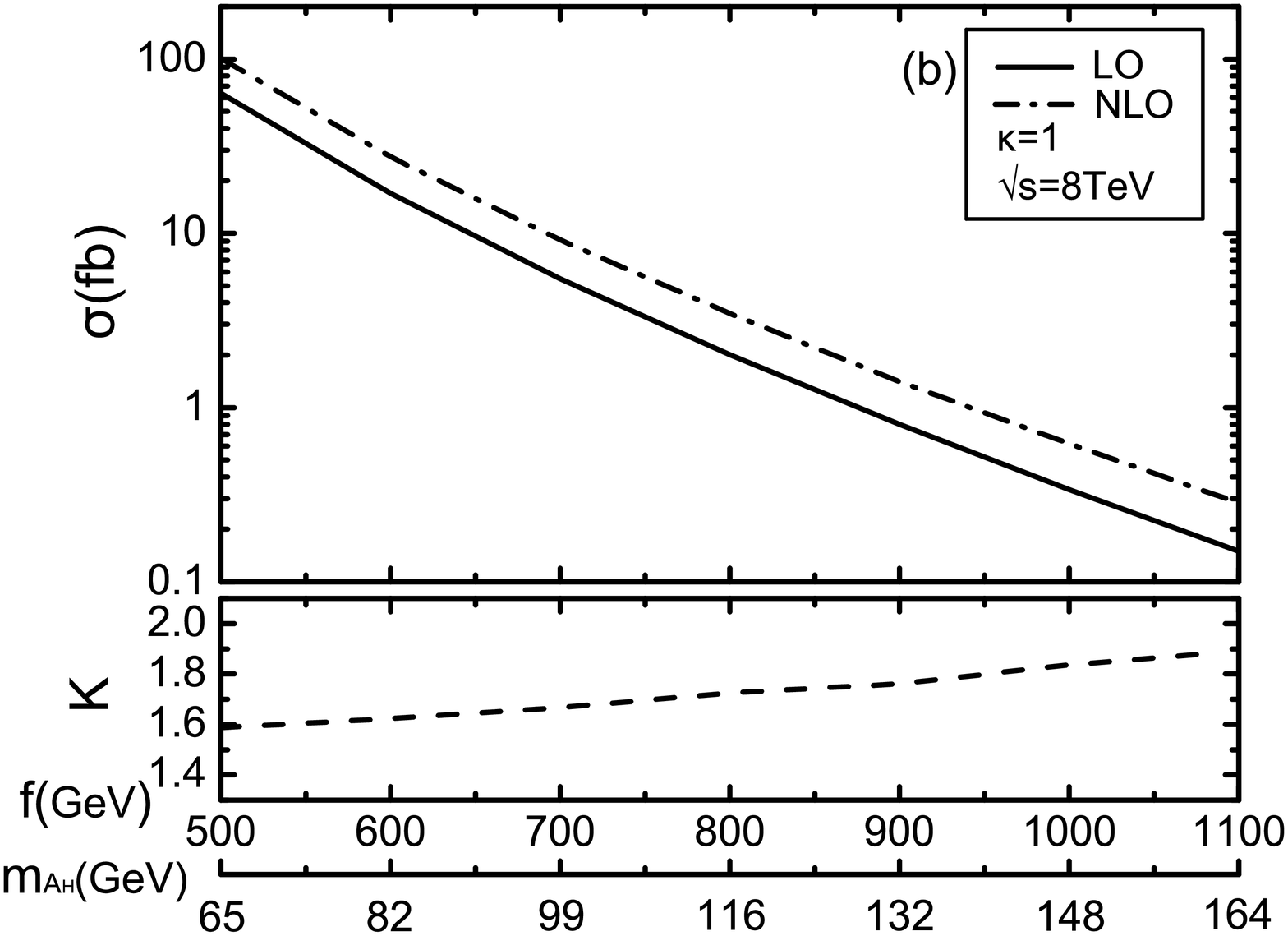}
\caption{ \label{fig7} The dependence of the LO, NLO QCD corrected
integrated cross sections and the corresponding $K$ factors on the
global symmetry-breaking scale $f$ for the $pp \to A_H q_- + X$
process at the LHC. (a)$\sqrt{s} = 14~{\rm TeV}$. (b) $\sqrt{s} =
8~{\rm TeV}$. }
\end{center}
\end{figure}

We plot the LO, NLO QCD corrected integrated cross sections and the
corresponding $K$ factors as functions of the global symmetry-breaking 
scale $f$ at the $\sqrt{s} = 14$ and $8~{\rm
TeV}$ LHC in Figs.\ref{fig7}(a) and \ref{fig7}(b), respectively.
There we take $\kappa = 1$ with $f$ varying from $500~{\rm GeV}$ to
$1.5~{\rm TeV}$ for the future LHC and $500~{\rm GeV}$ to $1.1~{\rm
TeV}$ for the present LHC. From the figures we find that the LO and
NLO QCD corrected cross sections for the $p p \to A_H q_- + X$
process decrease sensitively with the increment of $f$ because the
masses of final $A_H$ and $q_-$ become heavier, and the phase space
becomes smaller when $f$ goes up quantitatively. The numerical
results for some representative values of $f$ are presented in Table
\ref{tab4}.
%%%%%%%%%%%%%%%%%%%%%figure8%%%%%%%%%%%%%%%%%%%%%%%%%%%%%%%
%%%%%%%%%%%%%%%%%%%%%%table%%%%%%%%%%%%%%%%%%%%%%%%%%%%%%%%
\begin{table}
\begin{center}
\begin{tabular}{|c|c|c|c|c|}
\hline
$\sqrt{s}~ ({\rm TeV})$ &  ~$f~ ({\rm GeV})$~   & ~~$\sigma_{LO}~ (fb)$~~
& ~$\sigma_{NLO}~ (fb)$~ & ~~~$K$~~~   \\
\hline
%%% k=1, s=14
              & 500           & 346.51(3)      & 490(3)          & 1.41    \\
              & 700           & 43.164(3)      & 63.0(4)         & 1.46    \\
 14           & 900           & 9.0543(7)      & 13.64(8)        & 1.51    \\
              & 1100          & 2.4476(1)      & 3.84(2)         & 1.57    \\
              & 1300          & 0.77276(6)     & 1.245(7)        & 1.61    \\
              & 1500          & 0.27049(2)     & 0.456(2)        & 1.68    \\
\hline
              & 500           & 63.627(5)      & 100.6(7)        & 1.58    \\
 8            & 700           & 5.4902(4)      & 9.12(6)         & 1.66    \\
              & 900           & 0.80048(6)     & 1.408(8)        & 1.76    \\
              & 1100          & 0.14969(1)     & 0.283(1)        & 1.89    \\
\hline
\end{tabular}
\end{center}
\begin{center}
\begin{minipage}{15cm}
\caption{\label{tab4} The numerical results of $\sigma_{LO}$,
$\sigma_{NLO}$ and the corresponding $K$ factors for the $pp \to A_H
q_- + X$ process at the $\sqrt{s}=14$ and $8~{\rm TeV}$
LHC for some typical values of $f$ with $\kappa = 1$. }
\end{minipage}
\end{center}
\end{table}
%%%%%%%%%%%%%%%%%%%%%%%%%%%%%%%%%%%%%%%%%%%%%%%%%%%%%%%%%%%%%%%%%%%%%%%%%%%%%%%%%%%

\par
\subsection{Dependence on $T$-odd mirror quark Yukawa coupling $\kappa$ }
\par
\begin{figure}
\begin{center}
\includegraphics[width=0.48\textwidth]{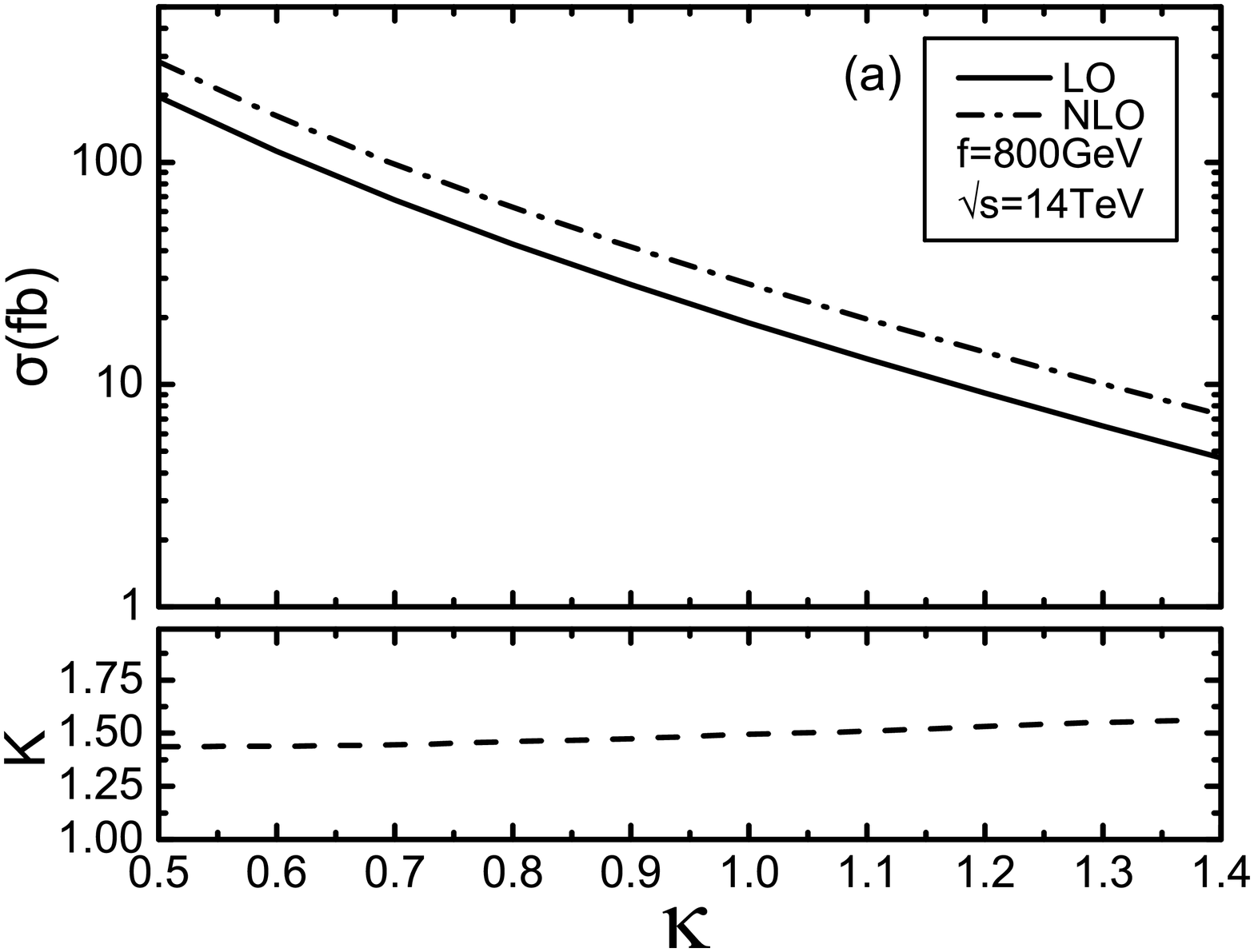}
\includegraphics[width=0.48\textwidth]{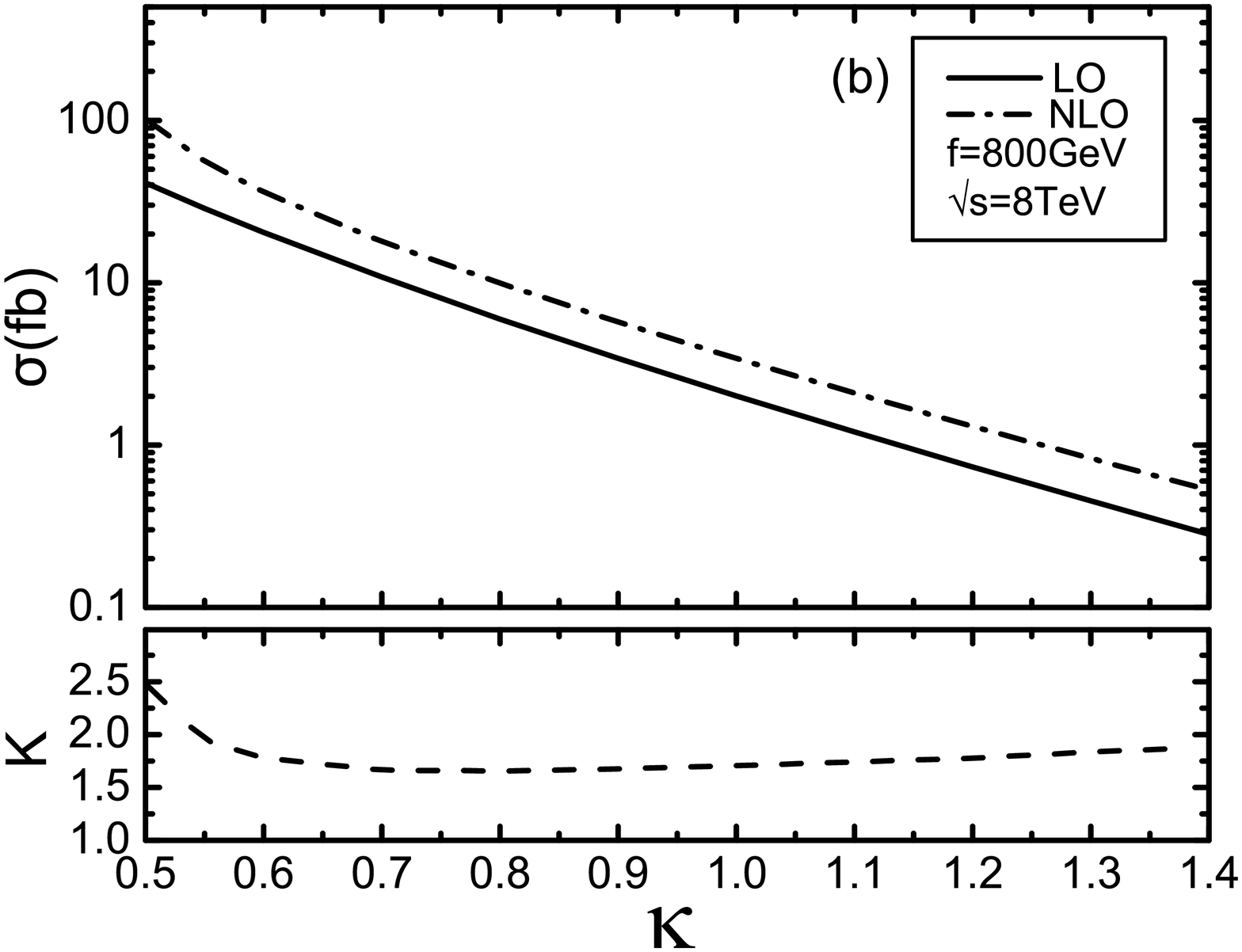}
\caption{ \label{fig8} The dependence of the LO, NLO QCD corrected
integrated cross sections and the corresponding $K$ factors on the
$T$-odd mirror quark Yukawa coupling $\kappa$ for the $pp \to A_H
q_- + X$ process at the LHC. (a) $\sqrt{s} = 14~{\rm TeV}$. (b)
$\sqrt{s} = 8~{\rm TeV}$. }
\end{center}
\end{figure}

The LO, NLO QCD corrected integrated cross sections and the
corresponding $K$ factors as functions of the $T$-odd mirror quark
Yukawa coupling $\kappa$ with $f =800~{\rm GeV}$ at the $\sqrt{s} =
14$ and $8~{\rm TeV}$ LHC are displayed in
Figs.\ref{fig8}(a) and \ref{fig8}(b), respectively. We can find
that the LO and NLO QCD corrected cross sections for the $pp \to A_H
q_- + X$ process at the LHC decrease with the increment of $\kappa$,
because the mass of final $T$-odd mirror quark $q_-$ is proportional
to $\kappa$. Some representative numerical results read off from
Figs.\ref{fig8}(a)-\ref{fig8}b) are listed in Table \ref{tab5}.
%%%%%%%%%%%%%%%%%%%%%figure8%%%%%%%%%%%%%%%%%%%%%%%%%%%%%%%%%
%%%%%%%%%%%%%%%Tablek%%%%%%%%%%%%%%%%%%%%%%%%%%%%%%tabelk%%%%%%%%%%%%%%%%%%%%%%%
\begin{table}
\begin{center}
\begin{tabular}{|c|c|c|c|c|}
\hline
$\sqrt{s}~ ({\rm TeV})$ &  ~~~$\kappa$~~~  &  ~~$\sigma_{LO}~ (fb)$~~  &  ~$\sigma_{NLO}~ (fb)$~ & ~~~$K$~~~  \\
\hline
               &   0.5          & 196.85(1)    & 282(7)          & 1.43  \\
               &   0.7          & 67.702(5)    & 97(1)           & 1.44  \\
       14      &   1.0          & 18.946(1)    & 28.3(3)         & 1.49  \\
               &   1.2          & 9.1341(7)    & 13.9(1)         & 1.53  \\
               &   1.4          & 4.6700(3)    & 7.30(7)         & 1.56  \\
\hline
               &   0.5          & 41.193(3)    & 102.0(4)        & 2.48  \\
               &   0.7          & 10.813(1)    & 17.9(2)         & 1.77  \\
       8       &   1.0          & 2.0106(1)    & 3.42(2)         & 1.71  \\
               &   1.2          & 0.73343(6)   & 0.828(8)        & 1.78  \\
               &   1.4          & 0.28175(2)   & 0.528(3)        & 1.88  \\
\hline
\end{tabular}
\end{center}
\begin{center}
\begin{minipage}{15cm}
\caption{\label{tab5} The numerical results of $\sigma_{LO}$,
$\sigma_{NLO}$ and the corresponding $K$ factors for the $pp \to A_H
q_- + X$ process at the $\sqrt{s}=14$ and $8~{\rm TeV}$
LHC for some typical values of $\kappa$ with $f = 800~{\rm GeV}$.}
\end{minipage}
\end{center}
\end{table}

\par
\subsection{Differential cross sections }
\par
In the LHT, $A_H$ is the lightest $T$-odd particle and therefore
stable, while $q_-$ is an unstable particle and mainly decays into
$W_H q^{\prime}$, $Z_H q$ and $A_H q$.
We assume the total decay width of $q_-$ to be the summation of
the partial decay widths for these three main decay channels, i.e.,
$\Gamma_{ q_-}\simeq \Gamma(q_- \to W_H q^{\prime}) + \Gamma(q_- \to
Z_H q) + \Gamma(q_- \to A_H q) $ ($q_- = u_-, c_-, d_-, s_-$),
and use the expressions of partial decay widths for the
$q_- \to W_H q^{\prime}$, $q_- \to Z_H q$ and $q_- \to A_H q$
decay channels presented in Ref.\cite{DuSM}. By taking
$f = 800~{\rm GeV}$ and $\kappa = 1$, we obtain
$Br(U_- \to W_H D(Z_H U))=60.32\%~(30.06\%)$, $Br(D_- \to
W_H U(Z_H D))=62.61\%~(31.39\%)$, $Br(U_- \to A_H U)=9.62\%$ and
$Br(D_- \to A_H D)=6.00\%$ where $U=u,d,~D=d,s$.

\par
For two-jet events (originating from the real emission corrections),
we apply the jet algorithm of Ref.\cite{jet} in the definition of
the tagged hard jet with $R=0.7$. That means when two jets in the
final state satisfy the constraint of $\sqrt{\Delta \eta^2 + \Delta
\phi^2} < R$, where $\Delta \eta$ and $\Delta \phi$ are the
differences of rapidity and azimuthal angle between the two jets, we
merge them into one new ``jet'' and consider the event as a one-jet
event with $p_{ij,\mu}=p_{i,\mu}+p_{j,\mu}$, and otherwise it
belongs to the two-jet event. We call the jet ($j_1$) with the
largest jet transverse energy ($E_T^{j_1}>E_T^{j_2}$) in two-jet
events the leading jet, while the jet in one-jet event is called the
leading jet too.

\par
\begin{figure}
\begin{center}
\includegraphics[width=0.48\textwidth]{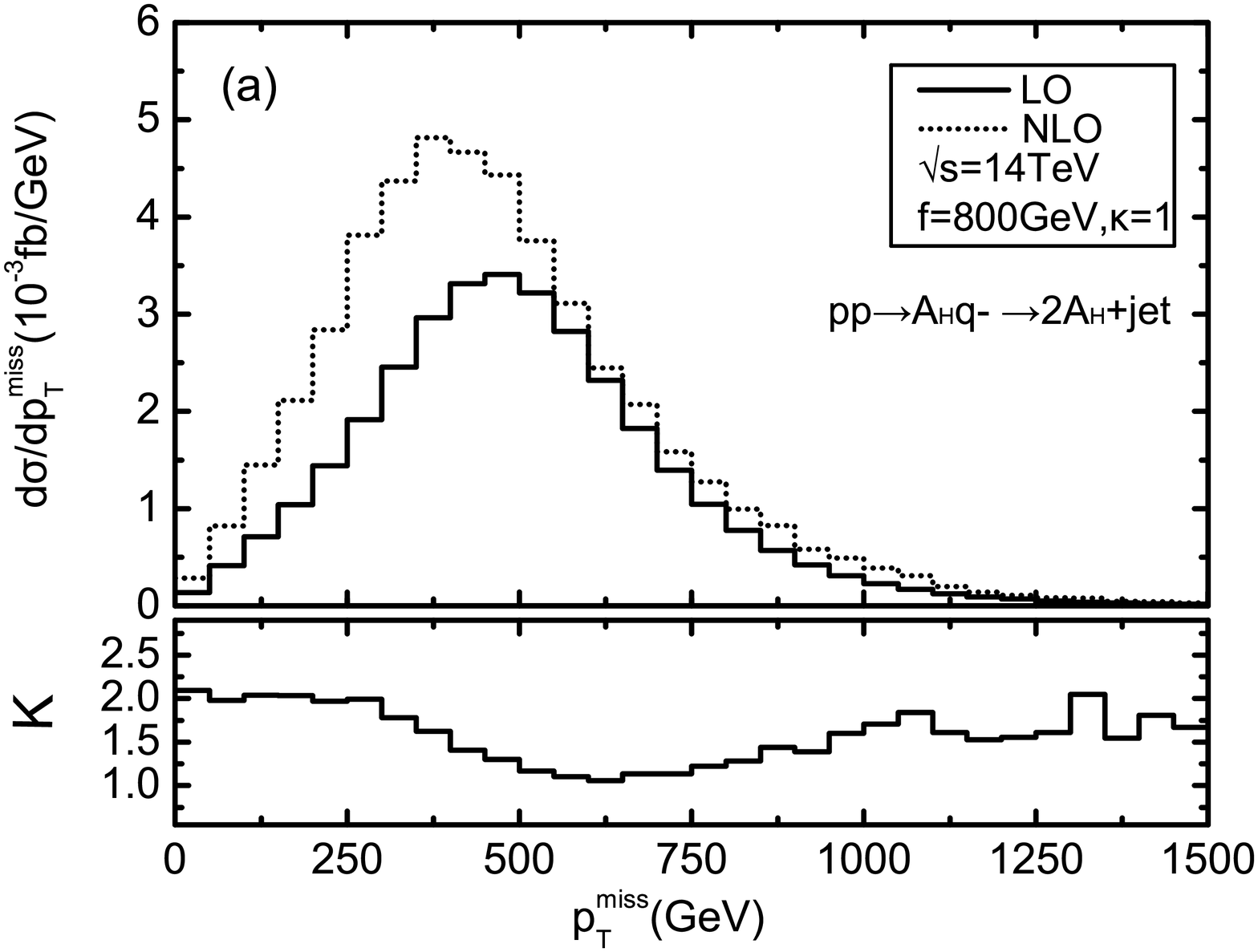}
\includegraphics[width=0.48\textwidth]{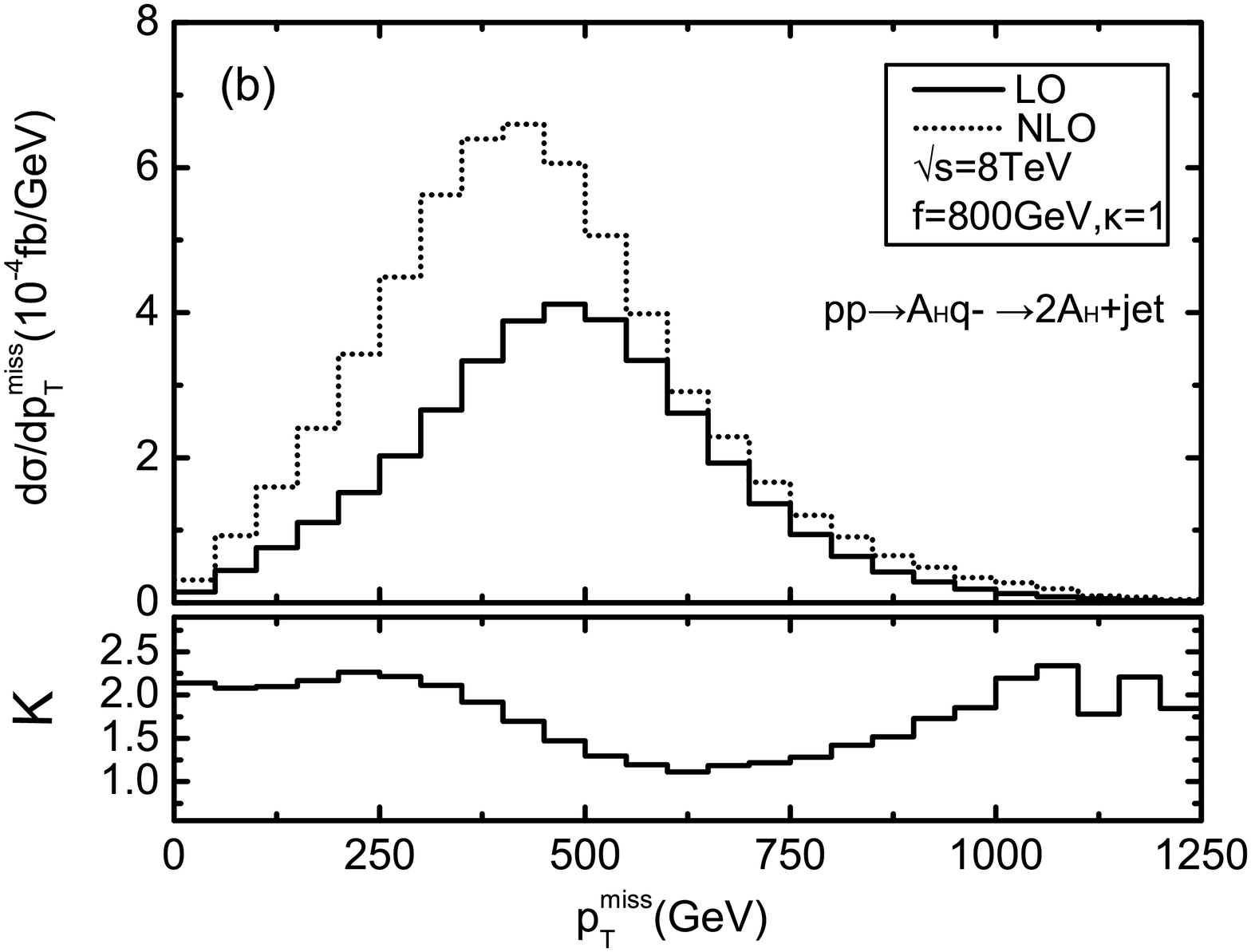}
\caption{ \label{fig9} The LO, NLO QCD corrected distributions of
the missing transverse momentum and the corresponding $K$ factors
for the $pp \to A_H q_- \to A_H A_H q + X$ process at the LHC. (a)
$\sqrt{s} = 14~{\rm TeV}$. (b) $\sqrt{s} = 8~{\rm TeV}$.}
\end{center}
\end{figure}

We first take the $A_H q_-$ production channel with the subsequent decay
$q_- \to A_H q$, i.e., $pp \to A_H q_- \to A_H A_H q + X$ process, as an
example to show the QCD NLO quantum effects on the LO differential cross sections.
Then a signal event of $A_H q_-$ associated production can be detected as
a single jet plus missing energy ($2 A_H$) in the LHC experiment.
We present the LO, NLO QCD corrected distributions of the missing
transverse momentum $p_T^{miss}$ and the corresponding $K$ factors
for the $pp \to A_H q_- \to A_H A_H q + X$ process at the $\sqrt{s}
= 14$ and $8~{\rm TeV}$ LHC in Figs.\ref{fig9}(a) and
\ref{fig9}(b), respectively. There we have $m_{A_H} = 115.64~{\rm GeV}$
by taking the LHT input parameters $f = 800~{\rm GeV}$ and $\kappa = 1$. We can see that
at both the $14$ and $8~ {\rm TeV}$ LHC, the LO $p_T^{miss}$ distributions
reach their maxima at the position of $p_T^{miss} \sim 475~{\rm GeV}$,
and the NLO QCD corrected $p_T^{miss}$ distributions have their maximal values
in the vicinity of $p_T^{miss} =400~{\rm GeV}$ at the present and future LHC.
Comparing the NLO corrected $p_T^{miss}$ distributions with the corresponding
LO ones in these two figures, we can see that the peak of the NLO QCD
corrected $p_T^{miss}$ distribution moves obviously to the left, which
would be due to the contributions of real gluon/light-quark emission
at the QCD NLO. The figures show that the corresponding $K$ factors
can exceed $2.0$ in the low $p_T^{miss}$ region, i.e.,
$p_T^{miss} < 250~ {\rm GeV}$ at the $\sqrt{s} = 14~ {\rm TeV}$ LHC
and $p_T^{miss} < 300~ {\rm GeV}$ at the $\sqrt{s} = 8~ {\rm TeV}$ LHC.
%%%%%%%%%%%%%%%%%%%%%%%%%%%%%%%%%%figure9%%%%%%%%%%%%%%%%%%%%%%%%%%%%%%

\par
\begin{figure}
\begin{center}
\includegraphics[width=0.48\textwidth]{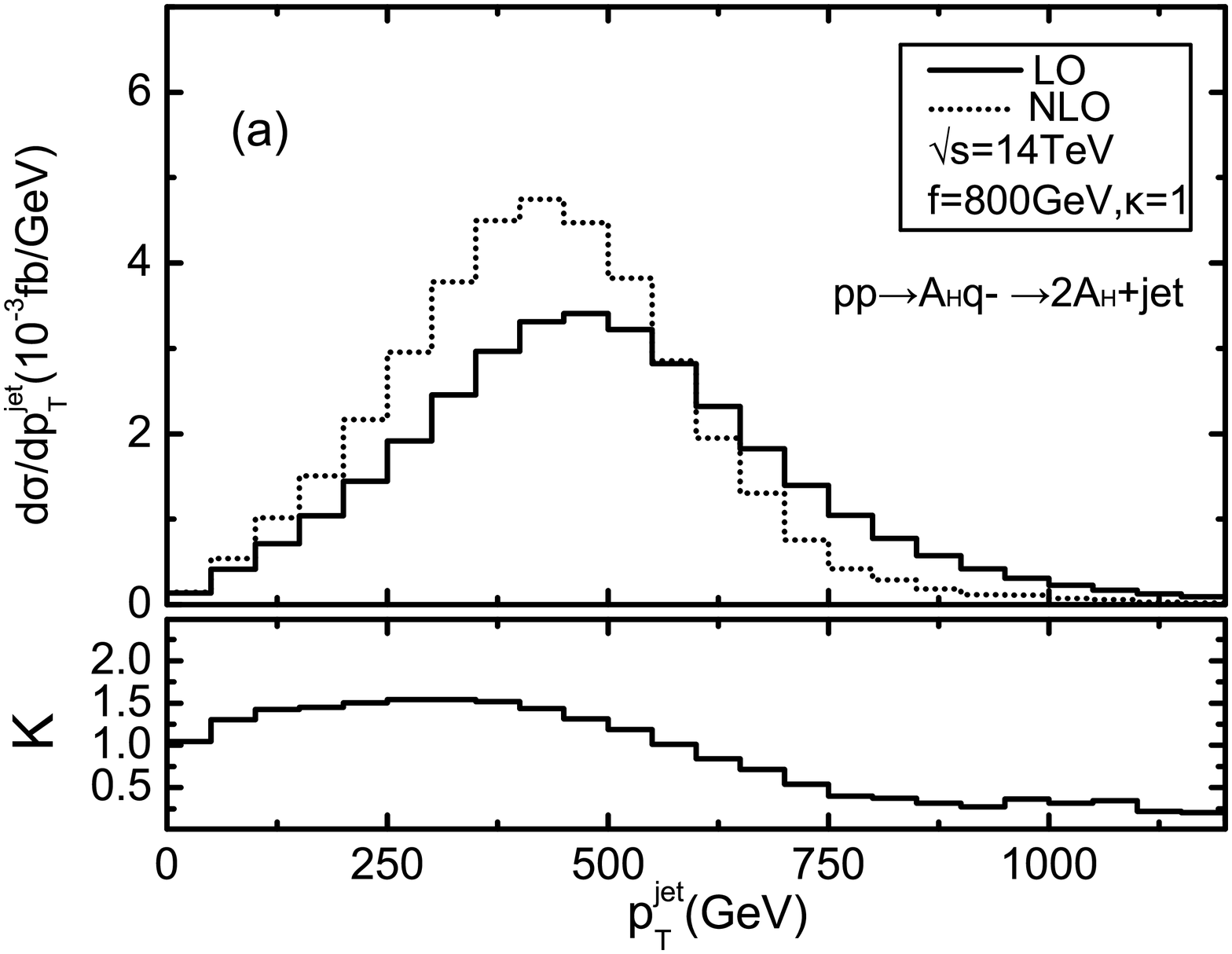}
\includegraphics[width=0.48\textwidth]{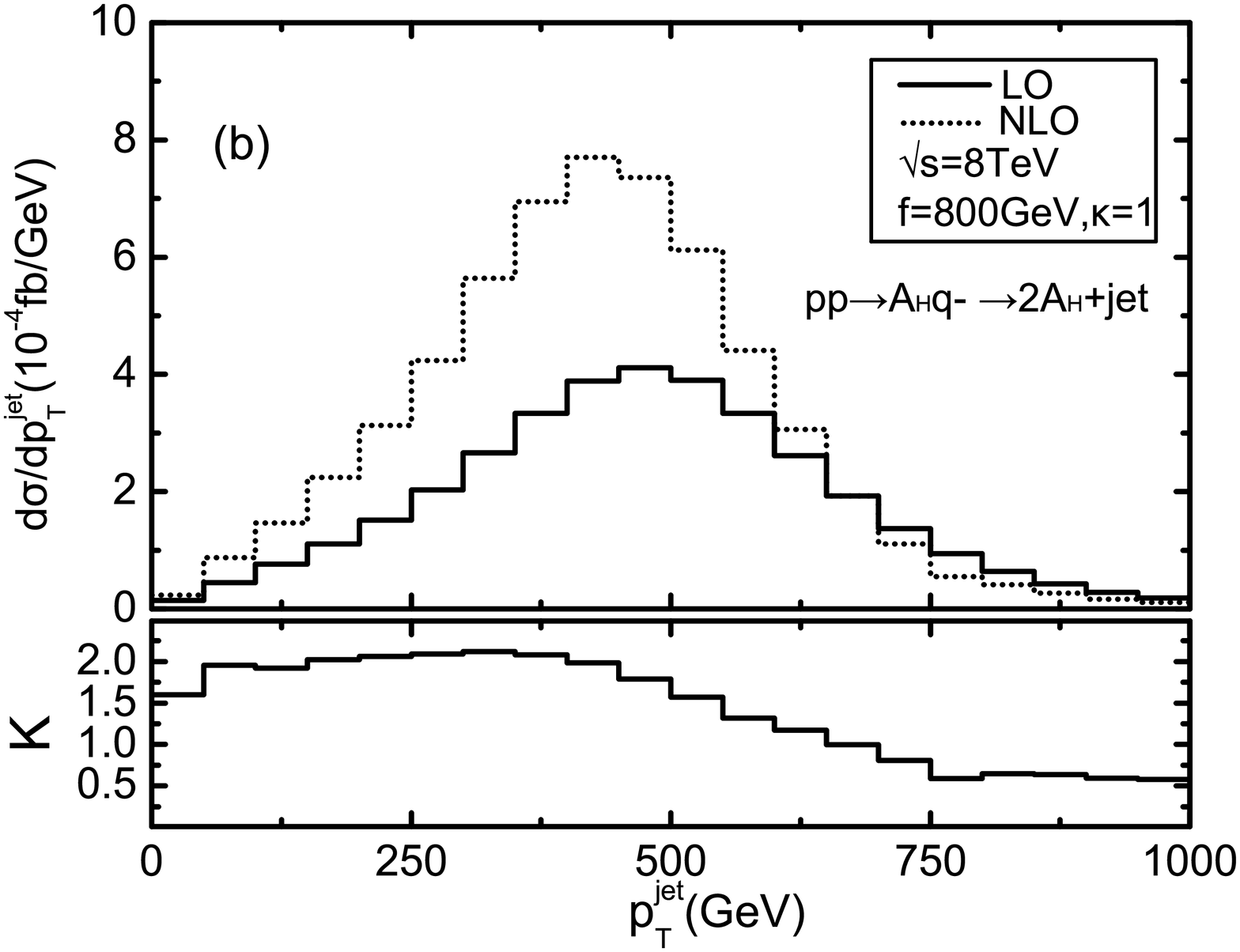}
\caption{ \label{fig10} The LO, NLO QCD corrected leading jet
transverse momentum distributions and the corresponding $K$ factors
for the $pp \to A_H q_- \to A_H A_H q + X$ process at the LHC. (a)
$\sqrt{s} = 14~ {\rm TeV}$. (b) $\sqrt{s} = 8~ {\rm TeV}$.}
\end{center}
\end{figure}

We depict the LO, NLO QCD corrected leading jet transverse momentum
distributions and the corresponding $K$ factors for the $pp \to A_H
q_- \to A_H A_H q + X$ process at the $\sqrt{s} = 14$ and
$8~{\rm TeV}$ LHC in Figs.\ref{fig10}(a) and \ref{fig10}(b), 
respectively. Both figures show that the LO and NLO QCD
corrected $p_T^{jet}$ distributions increase in the low $p_T^{jet}$
region and decrease in the high $p_T^{jet}$ region as the increment
of $p_T^{jet}$. And at both the future and present LHC the LO
$p_T^{jet}$ distributions reach their maxima in the vicinity of
$p_T^{jet} \sim 400~ {\rm GeV}$, while the NLO $p_T^{jet}$
distributions have their maxima at the position of $p_T^{jet} \sim
450~ {\rm GeV}$. We can also find that the NLO QCD corrections at
the $\sqrt{s}=14~{\rm TeV}$ ($8~{\rm TeV}$) LHC enhance the
$p_T^{jet}$ distributions in the range of $p_T^{jet} <
600~(700)~{\rm GeV}$, but reduce the $p_T^{jet}$ distributions in
the rest plotted $p_T^{jet}$ region.
%%%%%%%%%%%%%%%%%%%%%%%%%%%%%%%%%%figure10%%%%%%%%%%%%%%%%%%%%%%%%%%%%%%

\par
\begin{figure}
\begin{center}
\includegraphics[width=0.48\textwidth]{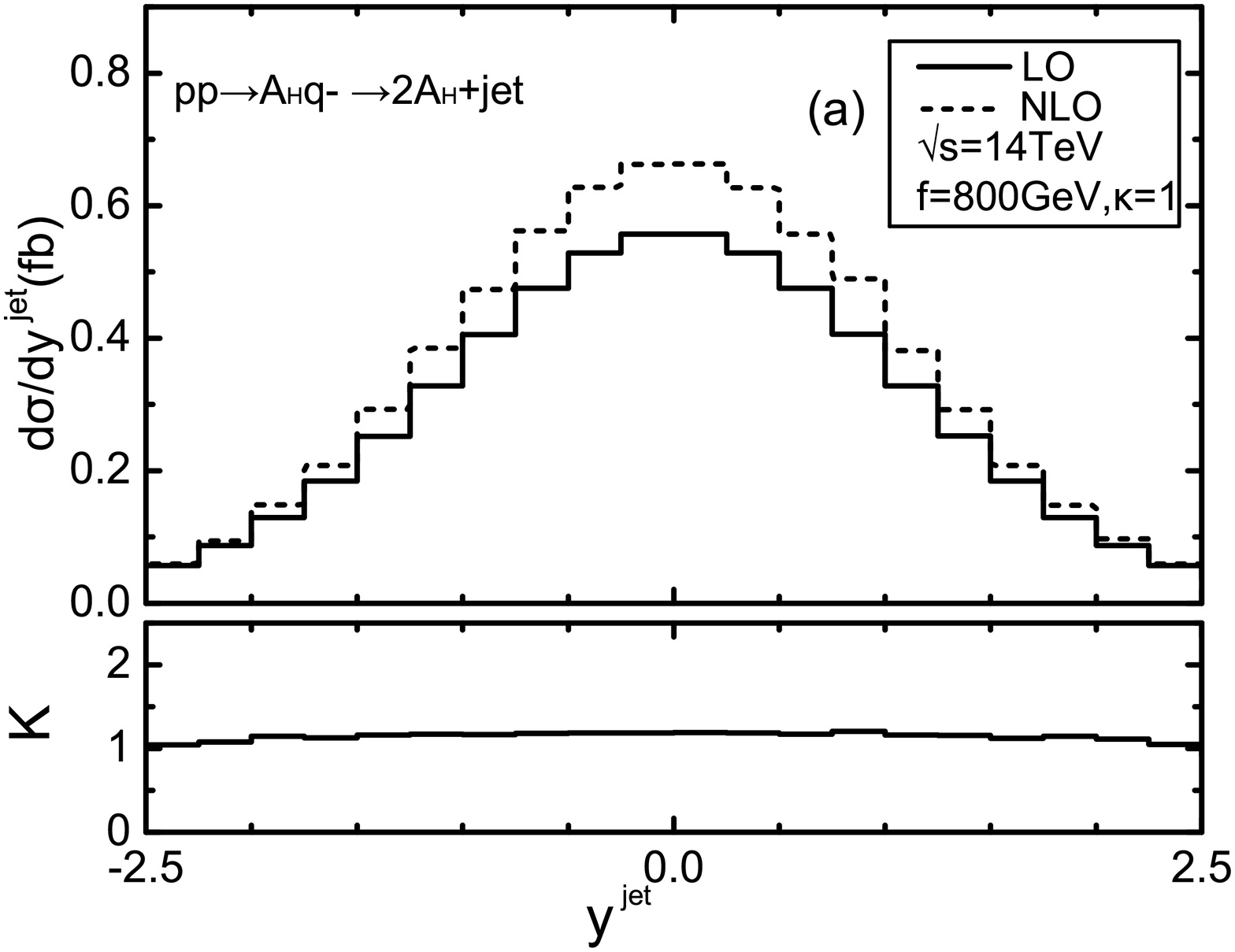}
\includegraphics[width=0.48\textwidth]{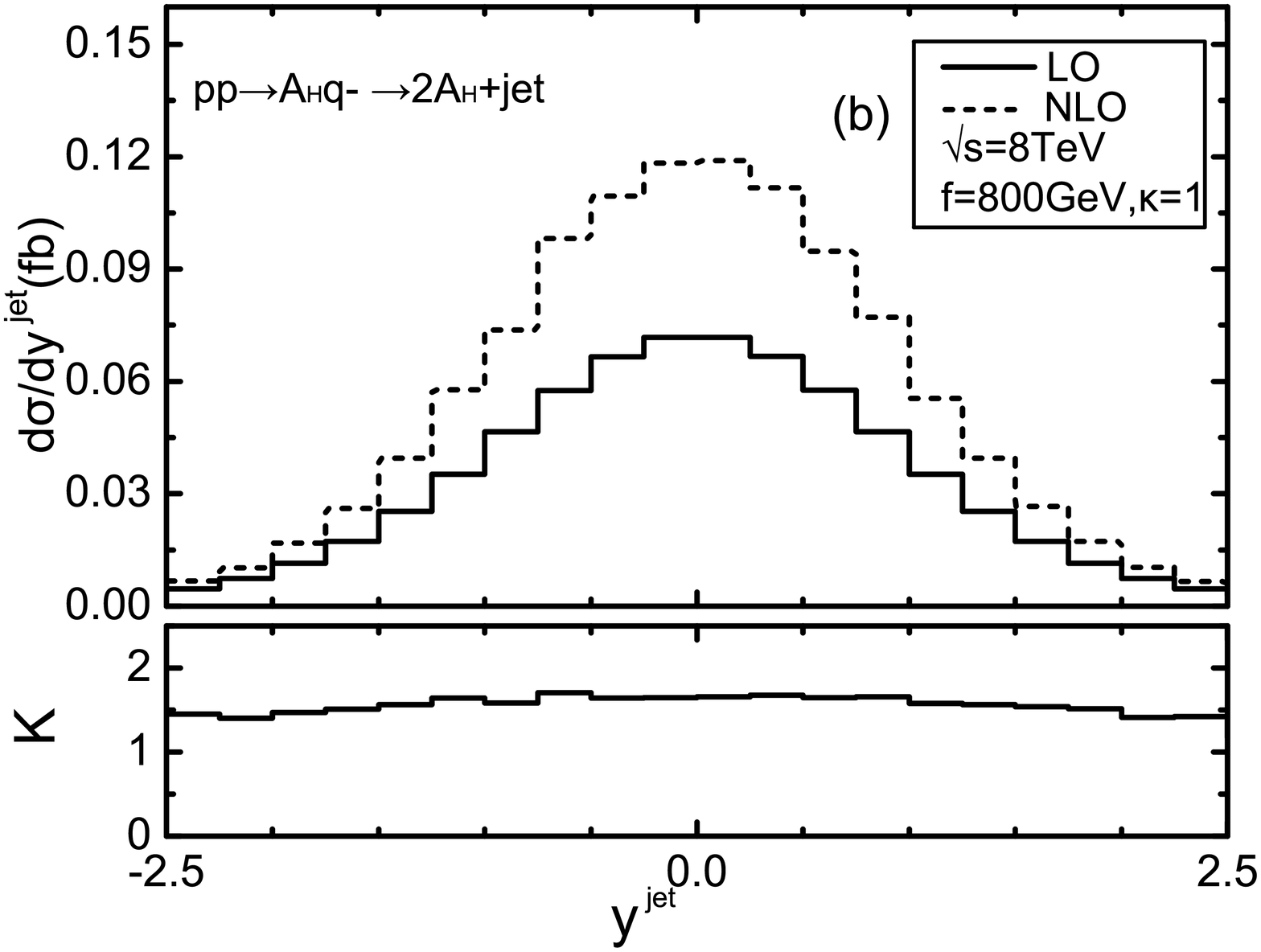}
\caption{ \label{fig11} The LO, NLO QCD corrected leading jet
rapidity distributions and the corresponding $K$ factors for the $pp
\to A_H q_- \to A_H A_H q + X$ process at the LHC. (a) $\sqrt{s} =
14~ {\rm TeV}$. (b) $\sqrt{s} = 8~ {\rm TeV}$.}
\end{center}
\end{figure}

The LO, NLO QCD corrected leading jet rapidity distributions and the
corresponding $K$ factors for the $pp \to A_H q_- \to A_H A_H q + X$
process at the $\sqrt{s} = 14$ and $8~{\rm TeV}$ LHC are
plotted in Figs.\ref{fig11}(a) and \ref{fig11}(b), respectively,
with $f = 800~{\rm GeV}$ and $\kappa = 1$. From Figs.\ref{fig11}(a)
and \ref{fig11}(b) we can see that the produced final jets are
mainly concentrated in the central rapidity region, and the
$K$-factor for the $y^{jet}$ distribution varies slightly in the
whole plotted $y^{jet}$ region.
%%%%%%%%%%%%%%%%%%%%%%%%%%%%%%%%%%figure11%%%%%%%%%%%%%%%%%%%%%%%%%%%%%%

\par
Recently the CMS Collaboration collected the events containing an
energetic jet and an imbalance in transverse momentum at the
$\sqrt{s}=8~{\rm TeV}$ LHC with an integrated luminosity of
$19.5~fb^{-1}$. It is found that the data are in good agreement with
expected contributions from SM processes \cite{CMS-Exp}. In Table 8
of Ref.\cite{CMS-Exp} the expected and observed $95\%$ C.L. upper
limits on possible contributions from new physics passing the
selection requirements are given. If we take the LHT signal as a
single jet event of the $pp \to A_H q_- \to 2 A_H q + X$ process.
Our calculation of the LHT signal process show that under the
present constraint on the scale $f$ the LHT signal does not have
significant impact on the present mono-jet event phenomenology. For
example, if we take $f = 700~{\rm GeV}$, $\kappa = 1$ and
$\mu=\mu_0$, we obtain less than three events of the LHT signal
passing the cut of $E^{miss}_T > 500~{\rm GeV}$ at the
$\sqrt{s}=8~{\rm TeV}$ LHC with integrated luminosity of
$19.5~fb^{-1}$, which is far below the $135$ events of the observed
$95\%$ C.L. upper limits on possible contributions from new physics
passing the selection cuts.

\par
\begin{figure}
\begin{center}
\includegraphics[width=0.48\textwidth]{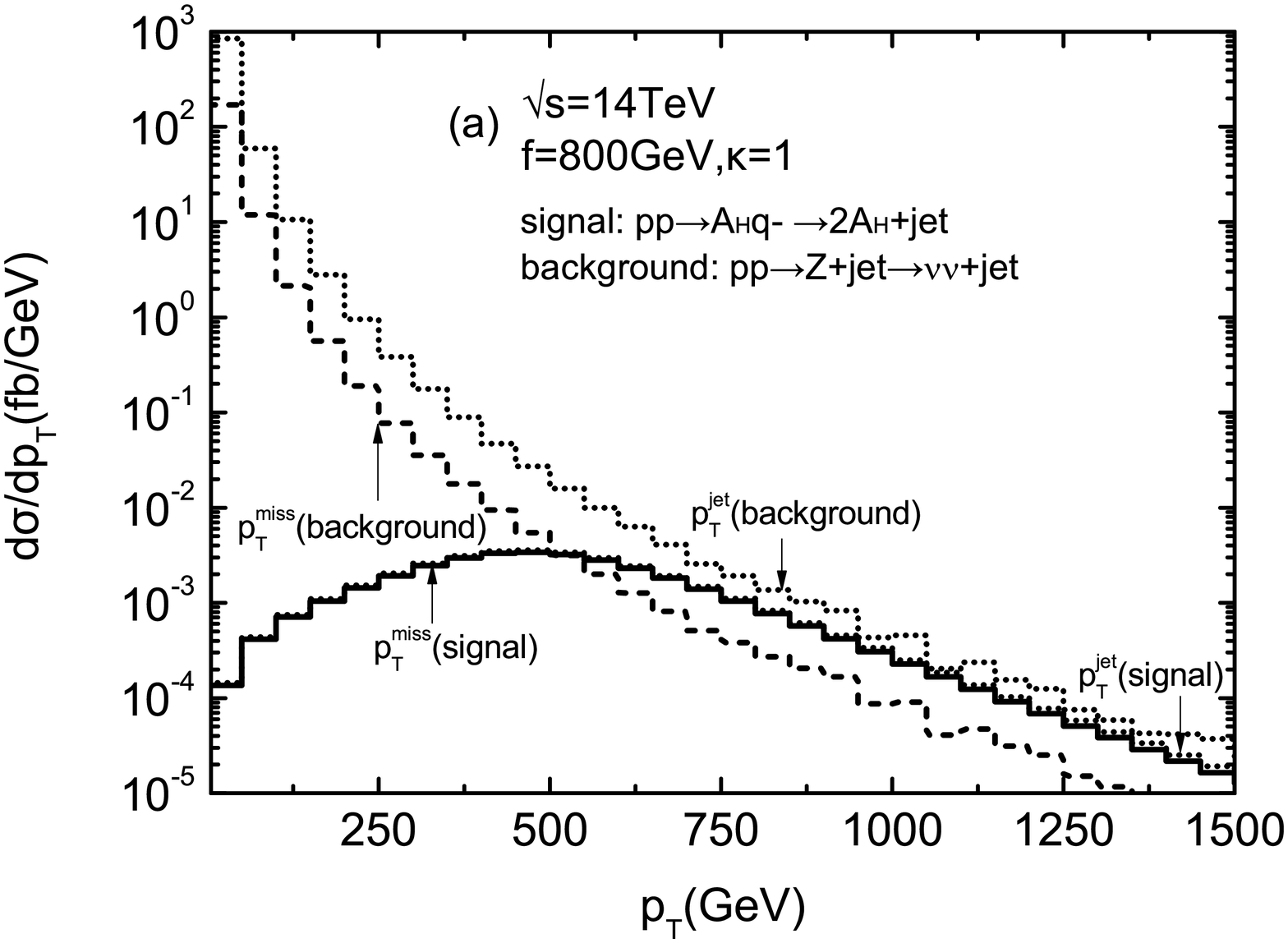}
\includegraphics[width=0.48\textwidth]{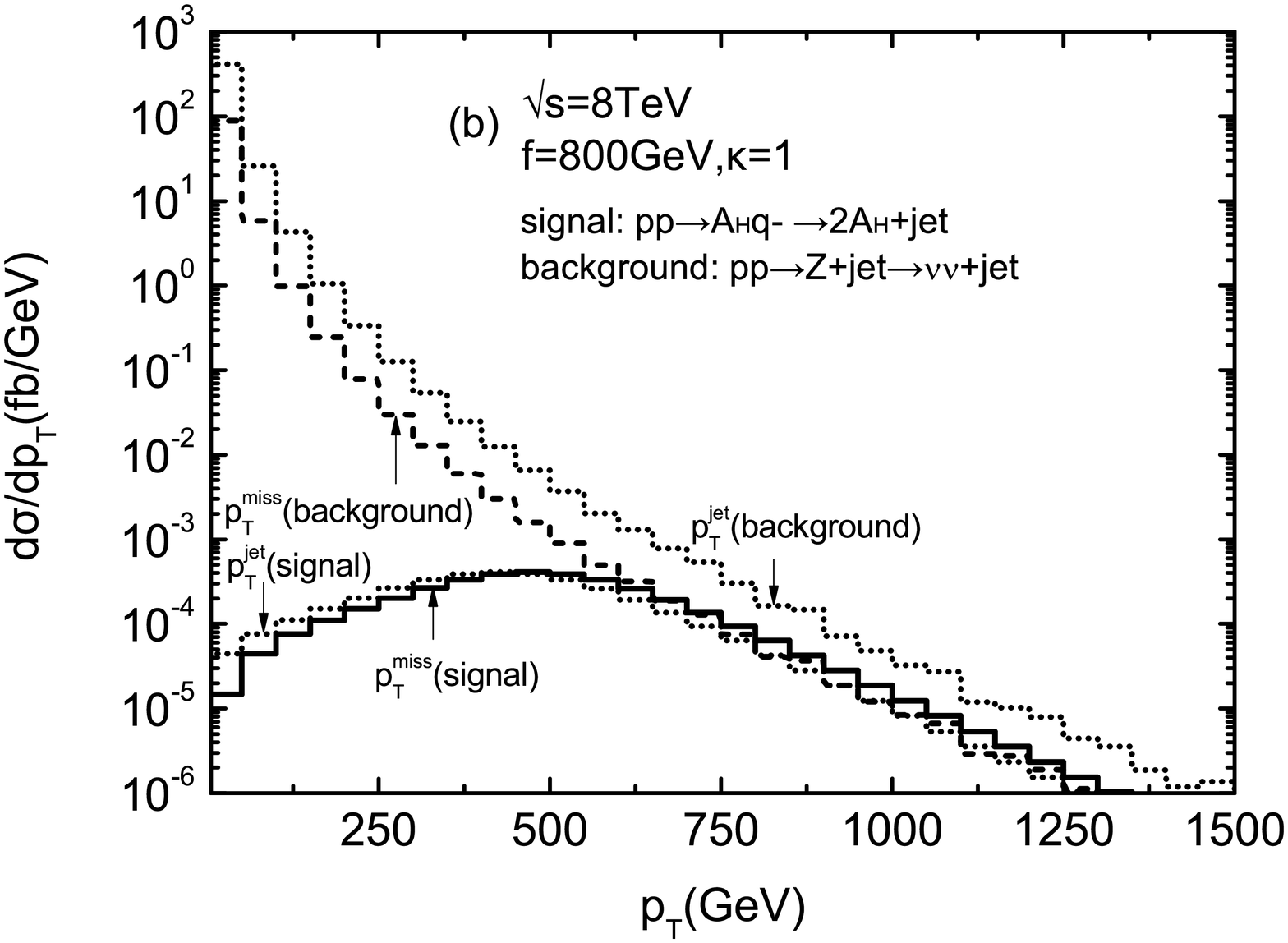}
\caption{ \label{fig12} The transverse momentum distributions for the signal process $pp \to A_H
q_- \to 2A_H q + X$ and the main SM background process $pp \to Z +
jet \to \nu \bar{\nu} + jet + X$ at the LHC. (a) $\sqrt{s} = 14~
{\rm TeV}$. (b) $\sqrt{s} = 8~ {\rm TeV}$.}
\end{center}
\end{figure}

In order to determine the event selection strategy in further data
analysis, we compare the kinematic distributions of the LHT signal
and SM background in the following discussion. When we choose
the LHT signal as the $A_Hq_-$ production followed
by the $q_- \to A_H q$ decay, i.e., $pp \to A_H q_- \to
2 A_H q + X$, its main SM background comes from the $pp \to Z + jet \to
\nu \bar{\nu} + jet + X$ process with single jet detected. We plot the
NLO QCD corrected distributions of the leading jet $p_T$ and the imbalance in transverse
momentum for the signal process $pp \to A_H q_- \to 2A_H q + X$, and the LO distributions
of $p_T^{jet}$ and $p_T^{miss}$ for the
main SM background process $pp \to Z + jet \to \nu \bar{\nu} + jet + X$
at the $\sqrt{s} = 14$ and $8~ {\rm TeV}$ LHC in
Figs.\ref{fig12}(a) and \ref{fig12}(b), respectively. From those two figures,
we can see that the final leading jet and undetectable particles of the SM background
process $pp \to Z + jet \to \nu \bar{\nu} + jet + X$ tend to concentrate in
the low $p_T$ region, while the final leading jet and undetectable particles
of the signal process prefer to be produced in the larger $p_T$ region
compared with those of the background process. Therefore, it is possible that the
SM background from the $pp \to Z + jet \to \nu \bar{\nu} + jet + X$ process
can be suppressed if we take appropriate transverse momentum
cuts on final leading jet or missing energy.
%%%%%%%%%%%%%%%%%%%%%%%%%%%%%%%%%%figure11%%%%%%%%%%%%%%%%%%%%%%%%%%%%%%

\par
\begin{figure}
\begin{center}
\includegraphics[width=0.48\textwidth]{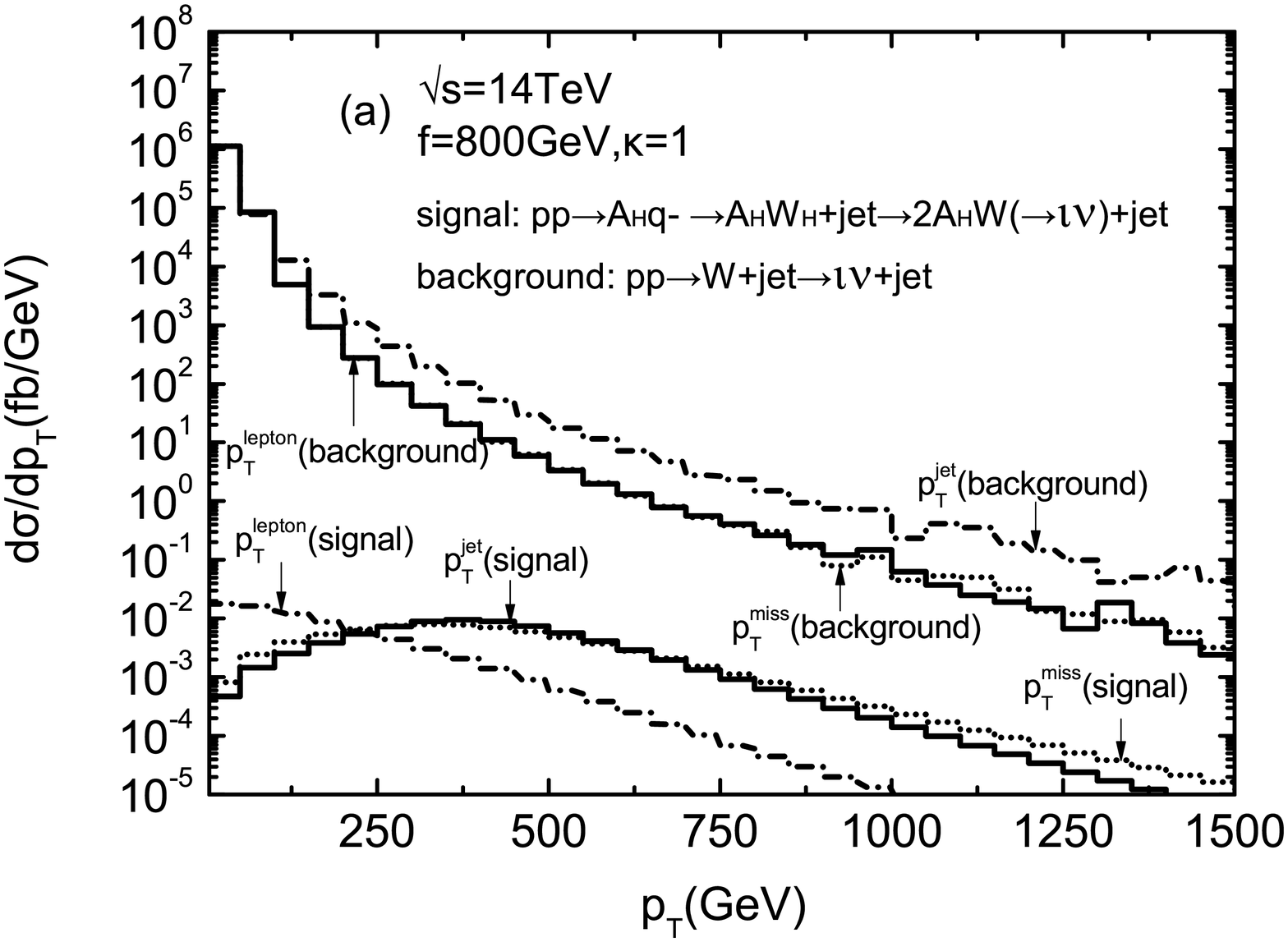}
\includegraphics[width=0.48\textwidth]{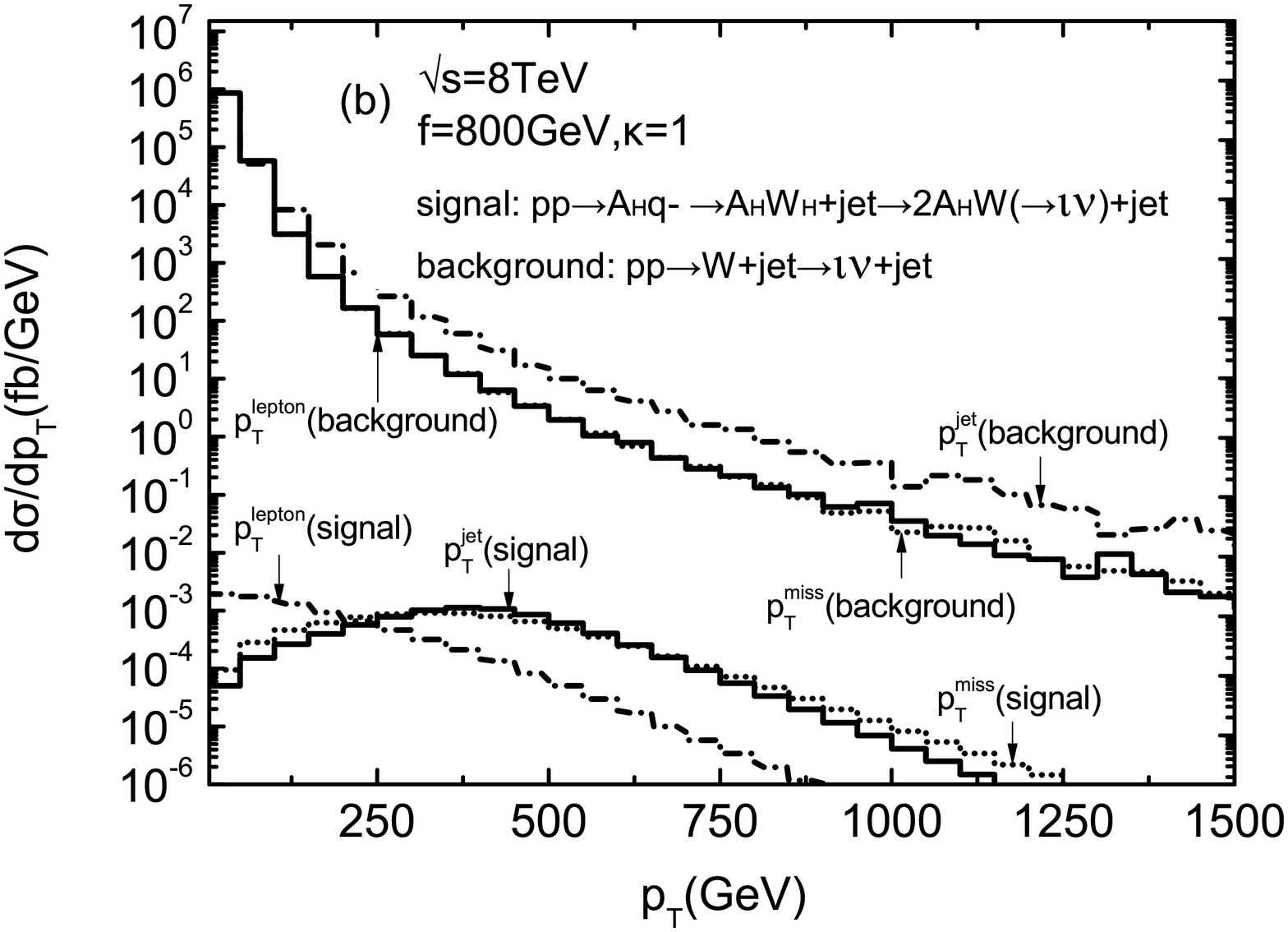}
\caption{ \label{fig13} The transverse momentum distributions for the signal process $pp \to A_H
q_- \to A_H W_H q' \to 2A_H W q' \to 2A_H l\nu q'+ X$ ($l=e,\mu,\tau$) and the
main SM background process $pp \to W + jet \to l \nu + jet + X$ ($l=e,\mu,\tau$)
at the LHC. (a) $\sqrt{s} = 14~ {\rm TeV}$. (b) $\sqrt{s} = 8~ {\rm
TeV}$.}
\end{center}
\end{figure}
\begin{figure}
\begin{center}
\includegraphics[width=0.48\textwidth]{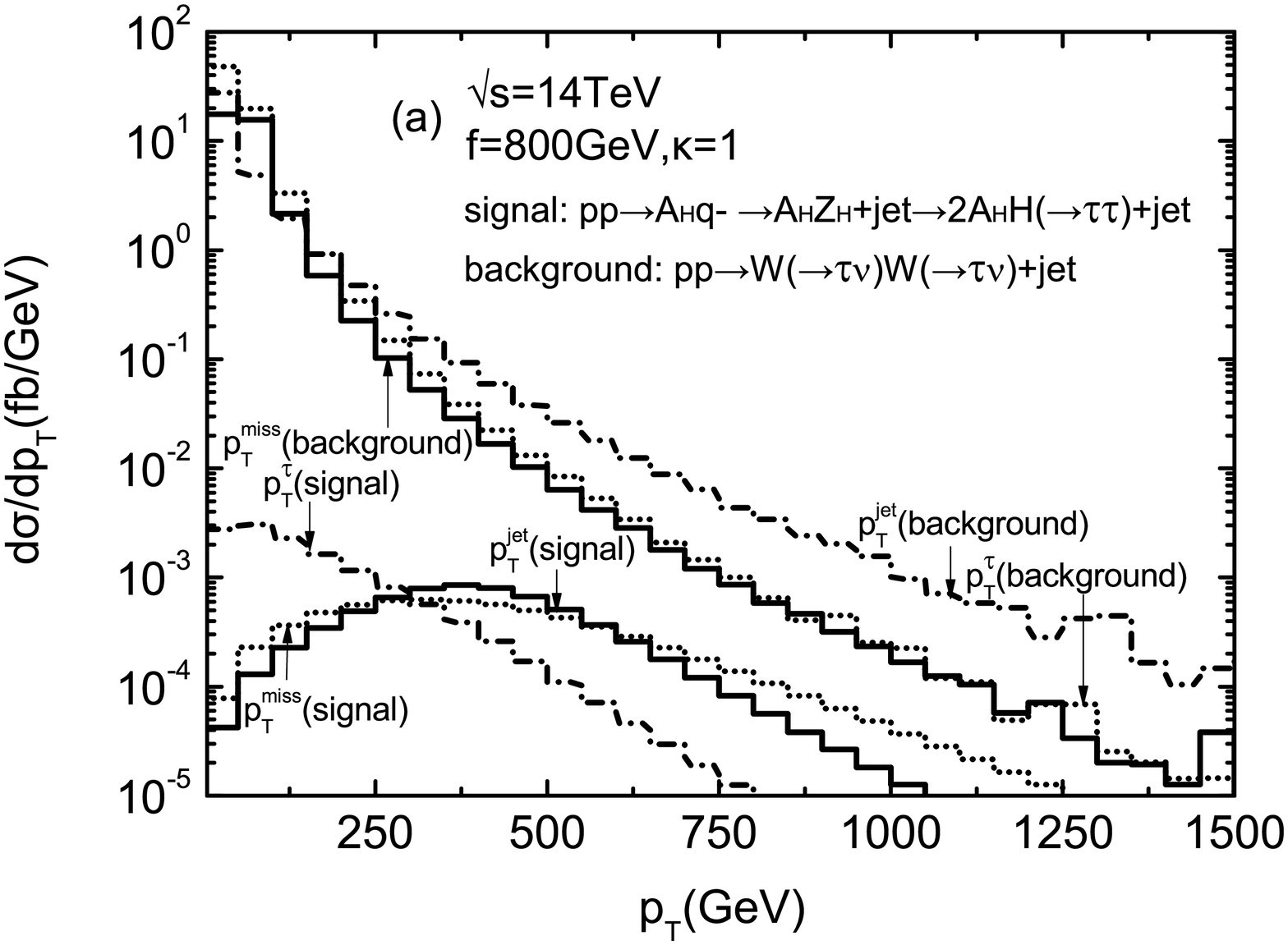}
\includegraphics[width=0.48\textwidth]{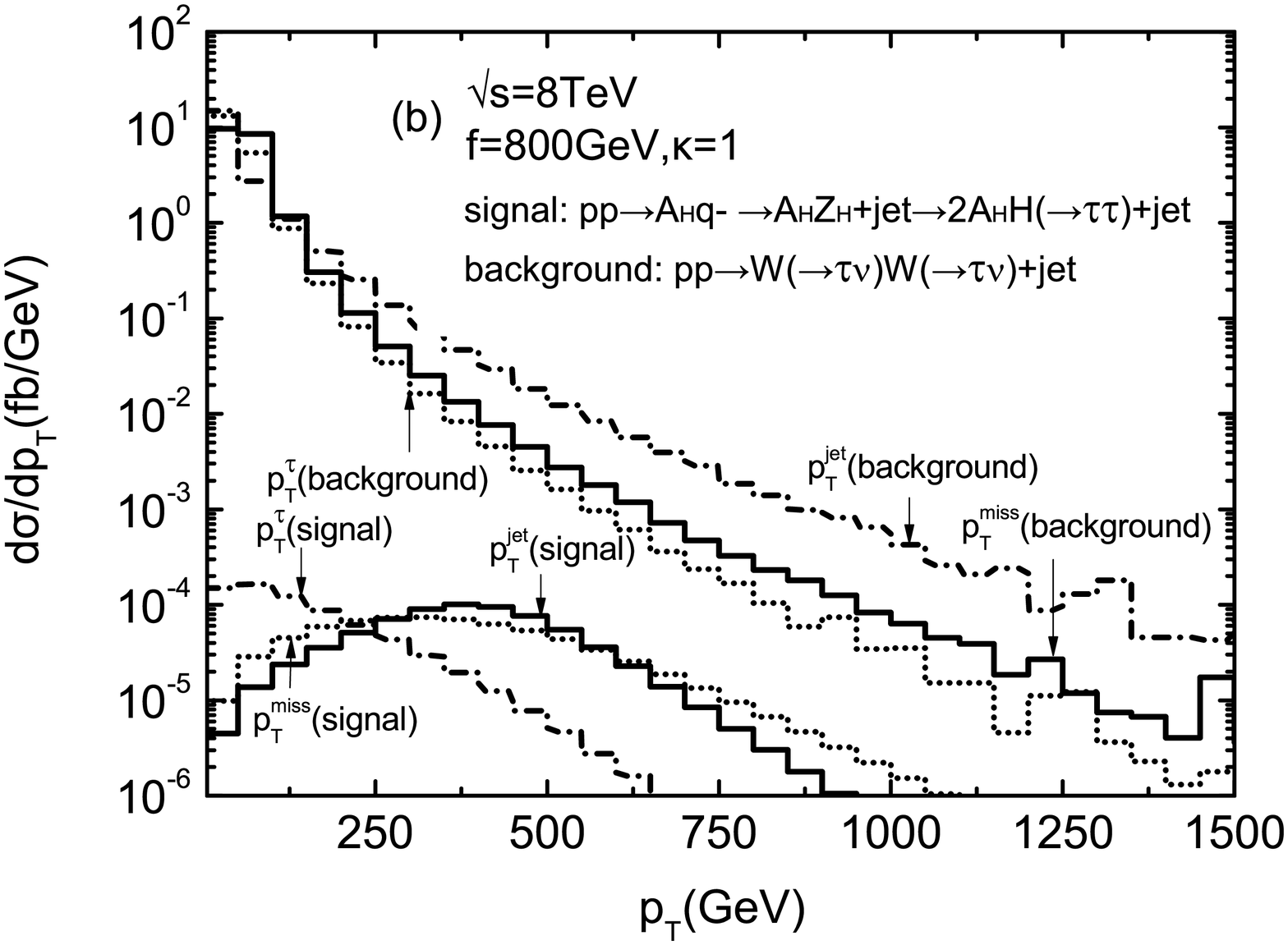}
\caption{ \label{fig14} The transverse momentum distributions for the
signal process $pp \to A_H q_- \to A_H Z_H q
\to A_H A_H H q \to 2A_H \tau^+\tau^- q + X$ and the main SM background
process $pp \to W^+W^- + jet \to  \nu \bar{\nu}\tau^+\tau^- + jet +
X$ at the LHC. (a) $\sqrt{s} = 14~ {\rm TeV}$. (b) $\sqrt{s} = 8~
{\rm TeV}$.}
\end{center}
\end{figure}

If we consider the processes $pp \to A_H q_- \to
A_H W_H q'\to 2 A_H W q^{\prime} \to 2A_H l\nu q'+ X$
($l=e,\mu,\tau$) and $pp \to A_H q_- \to A_H Z_H q
\to 2 A_H H q \to 2 A_H \tau^+\tau^- q + X$ as the LHT signals, these signals
could be detected as the lepton$+$jet$+$missing energy and $2\tau+$jet$+$missing energy
events, separately. The corresponding SM backgrounds mainly come from the
$pp \to W + jet \to l \nu + jet + X$ ($l=e,\mu,\tau$) and $pp \to W^+W^- +
jet \to \nu_{\tau} \bar{\nu}_{\tau} \tau^+\tau^- + jet + X$,
respectively. We plot the transverse momentum distributions of the final
leading jet, lepton and undetectable particles of the signal process
$pp \to A_H q_- \to 2A_H l\nu q+X$ and its main SM background process
$pp \to W + jet \to l \nu +jet + X$ in Fig.\ref{fig13}(a) for the
$14~ {\rm TeV}$ LHC and Fig.\ref{fig13}(b) for the $8~ {\rm TeV}$ LHC. 
And the signal process $pp \to A_H q_- \to 2 A_H \tau^+\tau^- q + X$
and the corresponding main SM background process $pp
\to W^+W^- + jet \to \nu_{\tau} \bar{\nu}_{\tau} \tau^+\tau^- + jet + X$ at the
$\sqrt{s} = 14~ {\rm TeV}$ and $8~ {\rm TeV}$ LHC are depicted in
Figs.\ref{fig14}(a) and \ref{fig14}(b), separately. From the
plots in Figs.\ref{fig13}(a)-\ref{fig13}(b) and Figs.\ref{fig14}(a)-\ref{fig14}(b), we can
find that the transverse momentum distributions of the final
leading jet and the undetectable particles of the signal are different
with those of background, because the kinematics of the signal is distinctively
different from that of background. The leading jet and the undetectable
particles in signal prefer to be distributed in the relatively large $p_T$
region except the final lepton and $\tau$, while the corresponding distributions of the
background events are concentrated in the low $p_T$ area. The $p_T$ distributions
of final lepton and $\tau$ for the signal processes in
Figs.\ref{fig13} and Figs.\ref{fig14} show a little special characteristic whose $p_T$ 
distribution shapes are similar with the corresponding ones for the SM backgrounds. From all
the six plots we can conclude that if we take some proper cuts on final
jet and missing energy, the SM background of the LHT signal could
be significantly suppressed.
%%%%%%%%%%%%%%%%%%%%%figure%%%%%%%%%%%%%%%%%%%%%%%%%%%
%%%%%%%%%%%%%%%%%%%%%figure%%%%%%%%%%%%%%%%%%%%%%%%%%%

\vskip 5mm
\section{Summary}
\par
In this paper, we calculate the $A_H q_-$ $(q_- = u_-, c_-, d_-,
s_-, \bar{u}_-, \bar{c}_-, \bar{d}_-, \bar{s}_-)$ associated
production rate at the $\sqrt{s} = 14$ and $8~ {\rm TeV}$
LHC up to the QCD NLO including the subsequent decay $q_- \to A_H q$
in the littlest Higgs model with $T$parity. We adopt the PROSPINO
strategy in real light-quark emission processes to avoid double
counting and provide reliable NLO QCD corrected predictions. We
investigate the dependence of the integrated cross section on the
factorization and renormalization scale $\mu$, the global symmetry
breaking scale $f$, and the $T$-odd mirror quark Yukawa coupling
$\kappa$. The distributions up to QCD NLO accuracy of the missing
transverse momentum, leading jet transverse momentum and rapidity
are also provided. Our numerical results show that the NLO QCD
correction enhances the LO integrated cross section remarkably with
the $K$ factor varying in the range of $1.41 \sim 1.68$ ($1.58 \sim
1.89$) as the increment of the global symmetry-breaking scale $f$
from $500~{\rm GeV}$ to $1.5~{\rm TeV}$ ($1.1~{\rm TeV}$) at the
$\sqrt{s} = 14~{\rm TeV}$ ($8~{\rm TeV}$) LHC. We also analyze
the distributions of the transverse momenta of final particles 
of the LHT signals and their SM backgrounds, and find that it is
possible to select the signal events of the $A_Hq_-$ production
from its background by taking proper cuts on the final leading 
jet and missing energy.
%%%%%%%%%%%%%%%%%%%%%%%%%%%%%%%%%%%%%%%%%%%%%%%%%%%%%

\vskip 5mm
\par
\noindent{\large\bf Acknowledgments:} This work was supported in
part by the National Natural Science Foundation of China
(Grants. No.11075150, No.11275190, No. 11375008), and the Fundamental
Research Funds for the Central Universities (Grant. No.WK2030040024).
X.-D. Yang would like to acknowledge support by the Fund for
Fostering Talents in Basic Science of the National Natural Science
Foundation of China (No.J1103207).

\end{document}